\documentclass[sigconf,screen]{acmart}
\AtBeginDocument{%
  }

\copyrightyear{2026}
\acmYear{2026}
\setcopyright{cc}
\setcctype{by-nc-nd}
\acmConference[SPAA '26]{38th ACM Symposium on Parallelism in Algorithms and Architectures}{July 06--10, 2026}{London, United Kingdom}
\acmBooktitle{38th ACM Symposium on Parallelism in Algorithms and Architectures (SPAA '26), July 06--10, 2026, London, United Kingdom}
\acmDOI{10.1145/3816782.3819219}
\acmISBN{979-8-4007-2761-0/2026/07}

\usepackage{amsmath}
\usepackage{amsthm}
\usepackage{amsfonts}
\usepackage{algorithm}
\usepackage[noend]{algpseudocode}
\usepackage{comment}
\usepackage{enumitem}
\usepackage{hyperref}
\usepackage{graphicx}
\usepackage{caption}
\usepackage{subcaption}
\usepackage{tcolorbox}
\usepackage{multicol}
\usepackage{subfiles}
\usepackage{xspace}
\usepackage{xcolor}
\usepackage{tablefootnote}
\usepackage{cleveref}

\newcommand{\myparagraph}[1]{\vspace{2pt} \noindent {\bf #1.}}

\DeclareMathOperator{\argmin}{argmin}
\DeclareMathOperator{\argmax}{argmax}

\newcommand{\calC}{\mathcal{C}}

\newcommand{\calL}{\mathcal{L}}
\newcommand{\calN}{\mathcal{N}}
\newcommand{\calP}{\mathcal{P}}

\newcommand{\calS}{\mathcal{S}}
\newcommand{\calT}{\mathcal{T}}
\newcommand{\calV}{\mathcal{V}}

\newboolean{full-version}
\setboolean{full-version}{true}

\newcommand{\sysname}[1]{{\textsc{#1}}}
\newcommand{\algoname}[1]{\textsc{#1}}
\newcommand{\varname}[1]{\textit{#1}}
\newcommand{\datasetname}[1]{{#1}}

\newcommand{\system}{\sysname{CleanANN}\xspace}

\newtheorem{definition}{Definition}
\newtheorem{routine}{Routine}

\newtheorem{challenge}{Challenge}

{\typeout{#1}}

\newcommand{\cosine}{\texttt{cosine}}

\usepackage[skip=2pt]{caption}

\begin{document}

\title{\system{}: Efficient and Robust Full Dynamism in Graph-based Approximate Nearest Neighbor Search}

\author{Ziyu Zhang}
\affiliation{%
  \institution{Massachusetts Institute of Technology}
  \city{Cambridge}
  \country{USA}
}
\email{sylziyuz@mit.edu}

\author{Yuanhao Wei}
\authornote{This work was done while at Massachusetts Institute of Technology.}
\affiliation{%
  \institution{University of British Columbia}
  \city{Vancouver}
  \country{Canada}
}
\email{yuanhaow@cs.ubc.ca}

\author{Joshua Engels}
\affiliation{%
  \institution{Massachusetts Institute of Technology}
  \city{Cambridge}
  \country{USA}
}
\email{jengels@mit.edu}

\author{Julian Shun}
\affiliation{%
  \institution{Massachusetts Institute of Technology}
  \city{Cambridge}
  \country{USA}
}
\email{jshun@mit.edu}


\begin{abstract}
 The approximate nearest neighbor search (ANNS) problem has important applications, such as robotics, data mining, semantic search, and unstructured data retrieval. Graph-based ANNS indexes have superb empirical tradeoffs in indexing cost, query efficiency, and query approximation quality. Most existing graph-based indexes are designed for the static scenario, where there are no updates to the data after the index is constructed. However, full dynamism (insertions, deletions, and searches) is crucial to providing up-to-date responses in applications using vector databases. 
    It is desirable that the index efficiently supports updates and search queries concurrently. Existing dynamic graph-based indexes have the following shortcomings: (1) graph repair operations incurred by delete queries involve many graph updates, hindering scalability; (2) data distribution shift and out-of-distribution queries are not addressed; and (3) in efficient systems implemented as directed graphs, clearing incoming edges after deletions is a global batch operation, delaying resource reuse and leading to periodic regressions in query quality.
    
    To solve these problems, we propose the \system{} system, which consists of three main components: (1) workload-aware linking of diverse search tree descendants to combat distribution shift; (2) query-adaptive on-the-fly neighborhood consolidation to efficiently handle deleted nodes; and (3) lock-free semi-lazy memory cleaning to accelerate the cleaning of stale information in the data structure and reduce the work spent by the first two components.
    
    We evaluate \system{} on three diverse datasets on fully-dynamic workloads and find that \system{} exhibits higher overall throughput while delivering similar query quality compared to previous state-of-the-art solutions, particularly in more challenging datasets with data distribution shift and out-of-distribution queries.
    On 64 threads, compared to the recent state-of-the-art system \sysname{Wolverine++}, \system{} delivers 12--18$\times$ overall throughput. Compared to the industry state-of-the-art \sysname{IP-DiskANN} system, \system{} delivers 1.14--1.26$\times$ overall throughput.
\end{abstract}



\acmYear{2026}\copyrightyear{2026}
\acmConference[SPAA '26]{38th ACM Symposium on Parallelism in Algorithms and Architectures}{July 6--10, 2026}{London, United Kingdom}
\acmBooktitle{38th ACM Symposium on Parallelism in Algorithms and Architectures (SPAA '26), July 6--10, 2026, London, United Kingdom}
\acmDOI{10.1145/3816782.3819219}
\acmISBN{979-8-4007-2761-0/26/07}


\maketitle

\section{Introduction}

$k$-nearest neighbor search 
is a well-studied problem with many applications where the relevance of data points can be captured by vector similarities, such as search engines~\cite{li2021embedding}, data clustering~\cite{chen2018fast}, and anomaly detection~\cite{shen2018anomaly}.
In these applications, datasets can be large and have several hundred dimensions, and need to be searched quickly and accurately.

\begin{definition}[$k$-Nearest Neighbors (kNN)]
Given a dataset $\datasetname{D}$ with $|\datasetname{D}|\gg k$ in a space $M$ with distance function $d(\cdot, \cdot)$, for a query $q\in M$, its set of $k$ nearest neighbors $\varname{kNN}(q)$ is a subset of $\datasetname{D}$ of size $k$ satisfying
$
    \forall x\in \datasetname{D}\setminus \varname{kNN}(q), \forall y\in \varname{kNN}(q), d(y, q) \leq d(x, q).
$
\end{definition}

Solving kNN exactly in high dimensions
is believed to be computationally hard~\cite{rubinstein_stoc18_ann_hardness}, and thus researchers focus on the approximate nearest neighbor search problem (ANNS) instead. This problem has become increasingly important due to the ubiquity of vector embedding-based semantic search for complex, multi-modal, unstructured data in AI applications \cite{lewis2021rag}. Data often change continuously in many applications that need ANNS, such as recommendation systems~\cite{suchal2010full}, autonomous agents~\cite{ocker2024tulip}, and robotics~\cite{plaku2008quantitative}. Dynamism in ANNS is important for data systems that support efficient data updates and semantic search, as this enables fresh responses to real-time inputs in AI applications.
This motivates the need for a \emph{dynamic} ANNS index that can keep up with, for example, the 500+ hours of content uploaded to YouTube every minute, the one billion images updated on JD.com every day~\cite{li2018design}, or the constant ingestion of emails and code by an AI assistant. We are interested in the realistic \emph{full dynamism} setting, where insertions, deletions, and searches can all happen concurrently.

Graph-based indexes such as \sysname{Vamana}~\cite{diskann} (the in-memory version of \sysname{DiskANN}), \sysname{HNSW}~\cite{my20_hnsw}, \sysname{NSG}~\cite{nsg}, and \sysname{NGT}~\cite{ngt_optimizations} have been shown~\cite{parlayann, ann_benchmark} to achieve state-of-the-art performance on static data (i.e., the index is not updated).\footnote{The best performing systems evaluated in ~\cite{ann_benchmark} are graph indexes with fine-grained optimizations such as quantization. These optimizations can be applied independently to the graph structure techniques that we focus on in this paper.}
However, existing dynamic graph-based indexes~\cite{freshdiskann, lsh_apg, hezel2023_deg, pgvector, xu2025inplaceupdatesgraphindex, liu2025wolverine} suffer from at least one of the following problems: (1) graph repair operations incurred by delete queries involve many graph updates, hindering scalability; (2) data distribution shift and out-of-distribution queries are not addressed; and (3) in efficient systems implemented as directed graphs, clearing incoming edges after deletions is a global batch operation, delaying memory reuse and leading to periodic regressions. We present these problems in detail in ~\Cref{subsection:limitations}.

We propose several techniques to address these issues. We propose \emph{bridge building}, a data-aware method that improves the robustness of graph-based indexes under full dynamism. The key idea is to add additional edges between visited points that are moderately far apart during graph traversals. In addition, we propose a \textit{dynamic neighborhood consolidation} method that identifies useful graph repair operations around deleted points and performs these repairs on the fly, without requiring global operations that degrade system throughput. We also propose \emph{semi-lazy cleaning}, a novel memory management strategy that efficiently cleans deleted nodes in the graph while maintaining graph quality, eliminating reliance on global batch cleaning operations and their associated regressions in query quality. 
We present \system, an ANNS index that achieves efficient full dynamism using these techniques. \system supports any mix of concurrent insertions, deletions, and searches.

\subsection{Limitations of Previous Work} \label{subsection:limitations}
\myparagraph{Graph-based Indexes} \label{subsubsection:graph_index_brief_intro}
In a graph-based ANNS index $G = (V, E)$ for dataset $\datasetname{D}$, each node $v_x\in V$ represents a data point $x\in \datasetname{D}$. Each graph-based index has its own algorithm for deciding how to connect edges among these nodes. Given a query point $q\in M$, the index generally conducts a \emph{greedy search} for $q$ on $G$ from a set of starting nodes, i.e., the search repeatedly explores the node closest to $q$ among unexplored neighbors in the search frontier. When there are no nodes among the unexplored neighbors closer than the best visited node, the best visited nodes are reported as the approximate $k$-nearest neighbors, or $\varname{aKNN(q)}$. The graph is usually constructed under a \textit{sparsity constraint}, typically expressed as an upper bound for the out-degree, to bound the complexity of the greedy searches. In practice, the de facto measure for the quality of $\varname{aKNN(q)}$ is the \textbf{recall}:\footnote{Usually "recall" means the proportion of correct data points returned, and "precision" means the proportion of correct data points among the ones returned. The measurement of interest here incorporates both and is colloquially referred to as recall.} 
\begin{definition}[Recall k@k]\label{definition:recall}
    Given a dataset $\datasetname{D}$ and a query $q$, let $\varname{aKNN}(q)\subseteq \datasetname{D}$ with $|\varname{aKNN}(q)| = k$ be the result set returned by an ANNS algorithm $\algoname{Alg}$. The recall k @ k of \algoname{Alg} for $q$ is
    $
        \frac 1 k|\varname{kNN}(q) \cap \varname{aKNN}(q)|.
    $
\end{definition}

Supporting dynamic operations, especially deletions, 
is a notoriously difficult problem for graph-based ANNS indexes.
Existing dynamic graph-based indexes~\cite{freshdiskann, hezel2023_deg, pgvector, xu2025inplaceupdatesgraphindex, liu2025wolverine, lsh_apg} suffer from at least one of the following problems:

\begin{challenge}[Deletion Efficiency Issue]
    Deletion algorithms involve a number of graph updates not bounded by the sparsity constraints of the graph, hindering performance.
\end{challenge}
\begin{challenge}[Robustness Issue]
    Deletion algorithms are local to the deleted points, and do not account for the insertion and search workloads or the overall graph structure. The insertion algorithms are not robust with respect to the order in which data points are inserted.
\end{challenge}
\begin{challenge}[Data Structure Staleness Issue]
    In efficient graph systems implemented as directed graphs, handling remaining incoming edges to deleted points relies on batch operations, delaying resource reuse and leading to periodic regressions in query quality.
\end{challenge}

\myparagraph{Deletion Efficiency Issue}
For each deleted point $p$, \sysname{IP-DiskANN} selects in-neighbors of $p$ encountered during a search for $p$. Since there is no upper bound on the in-degree of graph nodes (as opposed to the out-degree), the number of graph updates needed to repair deletions in these methods can be high. While \sysname{Wolverine++}~\cite{liu2025wolverine} bounds the number of in-neighbors updated to repair the graph in the case of a \algoname{Delete}, it may inspect the full 2-hop neighborhood of neighbors of $p$, resulting in expensive \algoname{Delete} queries.

\begin{figure}[t]
\centering
\includegraphics[width=\linewidth]{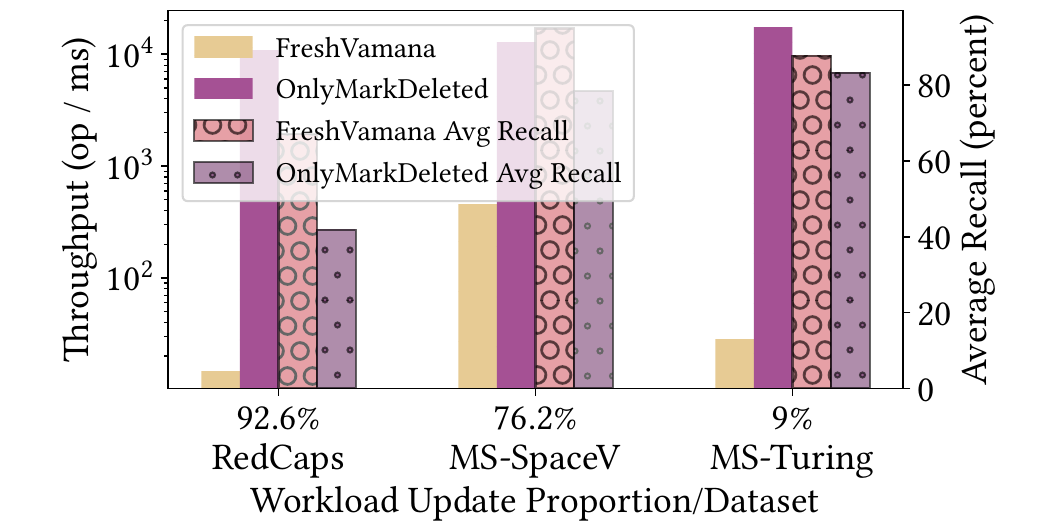}
\caption{Search throughput and average recall with (\sysname{FreshVamana}) or without (\sysname{OnlyMarkDeleted}) concurrent global consolidation. The datasets are described in \Cref{section:experiments}.}
\label{figure:batch_consolidate_slow}
\end{figure}

When data points are deleted, \sysname{FreshVamana}~\cite{freshdiskann} (the in-memory version of \sysname{FreshDiskANN}) first marks the corresponding nodes as \emph{tombstones} to filter them out in the responses for subsequent \algoname{Search} queries, and then periodically performs a global scan (known as \emph{consolidation}) across the entire index to update edges and remove tombstones. The consolidation algorithm tries to connect all in-neighbors of each tombstone to all out-neighbors of the same tombstone, which incurs a significant cost.
Our experiments show that the system throughput of \sysname{FreshVamana} can decrease by more than 10x
during consolidation; however, consolidation is required to achieve good \algoname{Search} recall (\Cref{figure:batch_consolidate_slow}). \sysname{LSH-APG}~\cite{lsh_apg} uses the same approach of connecting all in-neighbors to all out-neighbors to fix the graph after a node is deleted, implying similar efficiency problems. 

\begin{figure}[t]
\centering
\includegraphics[width=\linewidth]{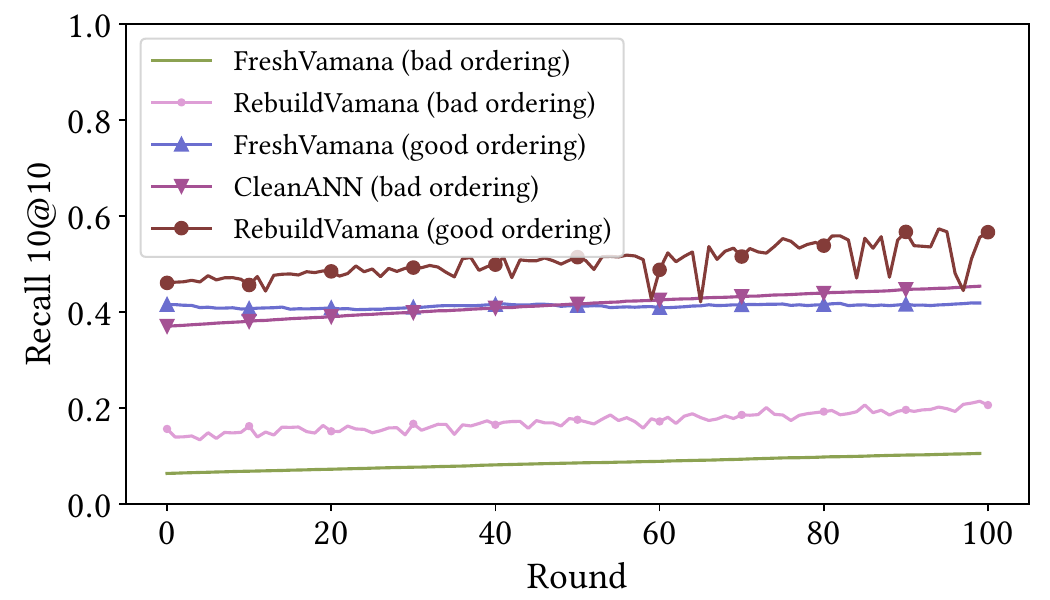} 
\caption{Synthetic dataset with a clustered distribution. A good ordering inserts whole clusters together and a bad ordering is uniformly random.}
\label{figure:insert_robustness_issue_motivation}
\end{figure}

\myparagraph{Robustness Issue}
Graph-based ANNS indexes are generally constructed by incrementally adding data points. A new data point $x\in\datasetname{D}$ is added via the following insertion routine:

\begin{routine}[New Data Point Insertion]
\label{routine:construction_and_ingestion}
\begin{enumerate}
    \item []
    \item Create a node $v_x\in V$ to represent $x$.
    \item Find a set of candidate neighbor nodes $\calC$ by conducting a best-first search for $x$.
    \item Choose good neighbors $\calC^{\star}\subseteq \calC$ with some index-specific heuristics and connect $v_x$ to $\calC^{\star}$.
\end{enumerate}
\end{routine}

For building the index, previous approaches combine ~\Cref{routine:construction_and_ingestion} with some additional optimizations. For example, \sysname{HNSW}~\cite{my20_hnsw} imposes a global hierarchical structure over the candidate set $\calC$ in step (2) and the edge choices in step (3); and \sysname{FANNG}~\cite{hd16_fanng} runs global iterative optimizations after running ~\Cref{routine:construction_and_ingestion} on each data point. Naturally, ~\Cref{routine:construction_and_ingestion} is used as an \algoname{Insert} algorithm for indexes. However, since the global optimizations mentioned above do not directly apply to the dynamic setting, \Cref{routine:construction_and_ingestion} when used as an \algoname{Insert} algorithm faces robustness issues.

Suppose a node $v_x$ has already been inserted.
For a node $v_y$ inserted later, 
let $\calC^{\star}_y$ be the nodes already in $G$ that are good neighbors for $v_y$. 
An edge from $v_x$ to $v_y$ is created only if $v_x \in \calC^{\star}_y$.
However, $v_y$ being a good neighbor for $v_x$ does not necessarily imply the converse, and an edge from $v_x$ to $v_y$ may never be created even if it is useful. Without global optimizations during the index build, \Cref{routine:construction_and_ingestion}, when used as an \algoname{Insert} algorithm, is sensitive to the ordering in which points are inserted and therefore not \emph{robust}. 

In~\Cref{figure:insert_robustness_issue_motivation}, 1\% of the dataset is inserted into the index on every round. We run a query batch after each insert batch, reporting the recall. \sysname{FreshVamana}, which directly uses ~\Cref{routine:construction_and_ingestion} to insert data points, shows a large recall difference with different orderings. Compared to \sysname{RebuildVamana}, which builds a \sysname{Vamana} index from scratch every round (a very inefficient algorithm in the dynamic setting), \sysname{FreshVamana} has about a $10\%$ lower recall under the same bad ordering, since it does not benefit from the global two-pass optimization of \sysname{RebuildVamana}. These comparisons show that~\Cref{routine:construction_and_ingestion} when used as an insertion algorithm is not robust.


\myparagraph{Data Structure Staleness Issue}
Since most performant graph indexes are implemented as adjacency lists that only store out-neighbors, given a deleted point $p$, one must scan the entire index to reliably identify all the in-neighbors of $p$. This is expensive, and thus existing methods (\sysname{IP-DiskANN} and \sysname{FreshDiskANN}) rely on periodic global batch operations to remove incoming edges to deleted points and free the corresponding resources. The amount of stale information (i.e., \textit{bloat}) accumulates over time and leads to query quality degradation until the bloat is removed by periodic cleaning operations, as we will see in~\Cref{section:experiments}.

\subsection{Our Contributions}
In this paper, we present the \system{} system with three main techniques that address the limitations discussed in \Cref{subsection:limitations}.
\begin{enumerate}
[topsep=1pt,itemsep=0pt,parsep=0pt,leftmargin=15pt]
    \item We propose \emph{bridge building}, a data-aware algorithm that improves the robustness of dynamic graph-based indexes. The key idea is to add additional edges between visited points that are moderately far apart during an insertion. This helps address the insertion robustness issue and allows the algorithm to achieve good recall without needing a good ordering in the workload. \Cref{figure:insert_robustness_issue_motivation} shows that \system with a bad ordering has the same average recall as \sysname{FreshVamana} with a good ordering.
    \item We propose a \emph{dynamic neighborhood consolidation} method that makes graph repairs around deleted points on the fly, avoiding global operations that degrade system throughput. The graph repair cost needed to actually delete a point is amortized across future operations.
    \item We propose \emph{semi-lazy cleaning}, a novel memory management strategy that efficiently cleans deleted nodes in the graph while maintaining graph quality. The key idea is to stop the consolidation and recycle the deleted nodes early to reduce work. Together with dynamic neighborhood consolidation, this addresses the deletion efficiency issue, enabling the system to achieve both high throughput and good recall. 
\end{enumerate}

Our techniques allow updates and queries to run concurrently. We evaluate \system{} on three datasets. Our main experiments focus on the sliding window setting. We use a batch-based design to measure recall under dynamism and concurrency following previous work \cite{diskann, spfresh, liu2025wolverine, xu2025inplaceupdatesgraphindex}. For each sliding window, we first issue an update batch that concurrently deletes old data points and inserts new data points into the index. Then, a batch of search queries runs
while the graph repairs incurred by the updates continue to run concurrently. We also present efficiency results from the same sliding windows but with inserts, deletes, and searches running concurrently. We compare against two recent state-of-the-art systems, \algoname{IP-DiskANN}~\citep{xu2025inplaceupdatesgraphindex} and \algoname{Wolverine++}~\cite{liu2025wolverine}. To ensure accurate comparison, we implemented \system{} and \algoname{Wolverine++}~\cite{liu2025wolverine} on the same base index as \algoname{IP-DiskANN}~\cite{xu2025inplaceupdatesgraphindex}.

Our experiments show that
\begin{enumerate}
[topsep=1pt,itemsep=0pt,parsep=0pt,leftmargin=15pt]
    \item \system{} efficiently supports both \algoname{Insert}s and \algoname{Delete}s without requiring expensive global operations or periodic rebuilds. In the full dynamism setting, \system{} achieves higher overall throughput across different datasets and workloads while maintaining similar or higher recall than two recent state-of-the-art baseline systems \sysname{Wolverine++}~\cite{liu2025wolverine} and \sysname{IP-DiskANN}~\cite{xu2025inplaceupdatesgraphindex}. On 64 threads, compared to \sysname{Wolverine++}~\cite{liu2025wolverine}, \system{} delivers 12--18$\times$ overall throughput. Compared to the industry state-of-the-art system \sysname{IP-DiskANN}~\cite{xu2025inplaceupdatesgraphindex}, \system{} delivers 1.14--1.26$\times$  overall throughput.
    \item \system{} maintains high recall and is robust against different insertion orderings and distribution shifts in the workload.
    \item \system{} scales well with respect to both dataset size and thread count, achieving 30--100 concurrent operations/ms on 64 threads.
\end{enumerate}


\section{Preliminaries}\label{section:preliminaries}
In graph-based ANNS indexes, each node corresponds to a data point. We use the letters $q$, $x$, and $y$ to denote data points and $u$, $v$, and $w$ to denote graph nodes. 
A subscript on a node (e.g., $v_x$) means that $v_x$ represents the data point $x$. Uppercase letters in the script typeface (e.g., $\calC$ and $\calN$) represent sets of graph nodes or edges. 
Subscripts on algorithm names (e.g., $\algoname{GreedyBeamSearch}_{L}$) denote parameters for the algorithms. \Cref{table:notations} lists more notation that we use in our algorithm descriptions.

\begin{table}[t]
\small
  \caption{Notation}
  \label{table:notations}
  \begin{tabular}{cl}
    \toprule
    Notation & Meaning \\
    \midrule
    $d(p, q)$ & Distance between points $p$ and $q$ \\
    $R$ & Graph out-degree bound \\
    $L$ & Backtracking budget for greedy search \\
    $L_I$ & Backtracking budget for greedy search during insertion\\
    $N(v)$ & Out-neighborhood of node $v$ \\
    $IN(v)$ & In-neighborhood of node $v$ \\
    $\alpha$ & Sparsity heuristic parameter \\
    $\pi(\cdot)$ & Parent of node in a tree\\
    $C$ & Eagerness threshold \\
  \bottomrule
\end{tabular}
\end{table}




The vanilla \algoname{GreedySearch} algorithm common in graph-based ANNS indexes may converge to a local optimum. Therefore, ANNS indexes commonly \emph{backtrack} to explore a few close-to-optimal unexplored nodes in the frontier to escape the local optimum. Our implementation is based on the \sysname{Vamana} graph. We refer readers to the \sysname{FreshDiskANN}~\cite{freshdiskann} paper for details of algorithms in \sysname{Vamana}: the backtracking \algoname{GreedySearch} (Algorithm 1,~\cite{freshdiskann}), \algoname{Insert} (Algorithm 2,~\cite{freshdiskann}), and the \algoname{RobustPrune} (Algorithm 3,~\cite{freshdiskann}) graph pruning algorithm in \sysname{Vamana}.

\section{Guided Bridge Building}\label{section:bridge}
In the static setting, previous work generally combines~\Cref{routine:construction_and_ingestion} with global optimizations to build the index. Existing methods~\cite{mplk14_nsw, my20_hnsw, hd16_fanng, nsg, diskann, freshdiskann, lsh_apg} use the same routine for insertions after building the index, including \sysname{FreshVamana}, which we implement our algorithms on, and \sysname{IP-DiskANN} and \sysname{Wolverine++} which we compare with. As discussed in~\Cref{subsection:limitations}, using~\Cref{routine:construction_and_ingestion} as \algoname{Insert} is not robust and exhibits search quality degradation compared to a static build.

Our algorithm \algoname{GuidedBridgeBuild} addresses this issue by adaptively introducing new edges based on the nodes traversed in \algoname{GreedySearch}. We will see that guided bridge building is critical to maintaining the quality of the index in the fully-dynamic setting in~\Cref{section:experiments}, improving both efficiency and quality for \algoname{Search} queries.

\subsection{Setup and Idea}\label{subsec:bridge_desc}
We explain \algoname{GuidedBridgeBuild} in the context of \sysname{FreshVamana}, which uses \algoname{GreedySearch} (Algorithm 1,~\cite{freshdiskann}) with \algoname{RobustPrune} (Algorithm 3,~\cite{freshdiskann}) under the framework of~\Cref{routine:construction_and_ingestion} to implement \algoname{Insert}. \algoname{GuidedBridgeBuild} selectively adds edges among the nodes visited by \algoname{GreedySearch}. Our method is orthogonal to the specifics of \algoname{RobustPrune} and backtracking in \sysname{FreshVamana} and may also be used to improve other graph-based indexes.

Let $\calV$ be the set of nodes explored by the search phase of \algoname{Insert}. $\calV$ can be viewed as a \emph{search tree} $\calT$, where $v$ is the parent of $w$ (denoted as $v=\pi(w)$) if $w$ was added to the search \emph{frontier} during the exploration of $v$, i.e., $w\in N(v)$ and $w$ had not been visited when $v$ was explored.
To insert $q$, existing methods~\cite{my20_hnsw, freshdiskann, nsg} connect $q$ to a subset of $\calV$. \sysname{FreshVamana} chooses this subset from $\calV$ with \algoname{RobustPrune}, which connects $q$ with nodes close to it in diverse directions to improve search quality while promoting shortcut edges to improve efficiency.

Our key observation is that creating bridge edges among \textit{cousins} in moderately deep levels (younger generations) of the search tree during a fully-dynamic workload creates paths that help with the robustness issue in \algoname{Insert}, 
similar to what static ANNS indexes achieve with global operations during the index construction phase.

Deeper-level cousins in the same search tree are likely to be nodes relatively close but not connected to each other via a short path, which can be caused by the robustness problem described in~\Cref{subsection:limitations}. As these cousin nodes are spatially close to each other and to the query, subsequent queries will likely need to navigate among these vertices. 
Without direct connections between these nodes, search queries require a larger backtracking budget to navigate between them, possibly resulting in early pruning of the search, which leads to lower search quality.

\algoname{GuidedBridgeBuild} introduces these missing connections. In addition to benefits in the dynamic case, we also find that \algoname{GuidedBridgeBuild} can improve the recall of static ANNS indexes by improving their robustness against bad
orderings during index construction.

\subsection{Algorithm Description}

\algoname{GuidedBridgeBuild} identifies good bridge edges via an augmented \algoname{GreedySearch}: \algoname{BridgeBuilderSearch} (\Cref{algorithm:guided_bridge_building}). Lines~\ref{line:bbbs_init_start}--\ref{line:bbbs_init_end} initialize the search frontier, the set of explored nodes, and the search tree $\calT$. 
Each iteration of Lines~\ref{line:bbbs_while_loop}--\ref{line:bbbs_explore_end} is a search step exploring the best unexplored node $w$.
On Line~\ref{line:bbbs_record_tree}, the algorithm marks unvisited out-neighbors of $w$ added to the frontier in the current iteration as the children of $w$ in $\calT$. Line~\ref{line:bbbs_call_gbb} invokes \algoname{GuidedBridgeBuild} to add the bridge edges based on $\calT$.
Nodes with search tree depths specified in $\calS$ are tentatively bi-directionally connected (Lines~\ref{line:gbb_candidate_pairs}--\ref{line:gbb_candidate_pairs_2}), subject to additional heuristic constraints by a commutative predicate \algoname{HeuristicPredicate} (Line~\ref{line:gbb_heuristic_predicate}).
Lastly, Line~\ref{line:gbb_add_neighbors} finalizes the bridge edges and performs additional pruning if needed via \Cref{algorithm:add_neighbors}.

\algoname{GuidedBridgeBuild} applies to both \algoname{Insert} and \algoname{Search} queries.

\begin{figure*}
    \centering
    \begin{subfigure}[b]{0.245\linewidth}
        \centering
        \includegraphics[width=\linewidth]{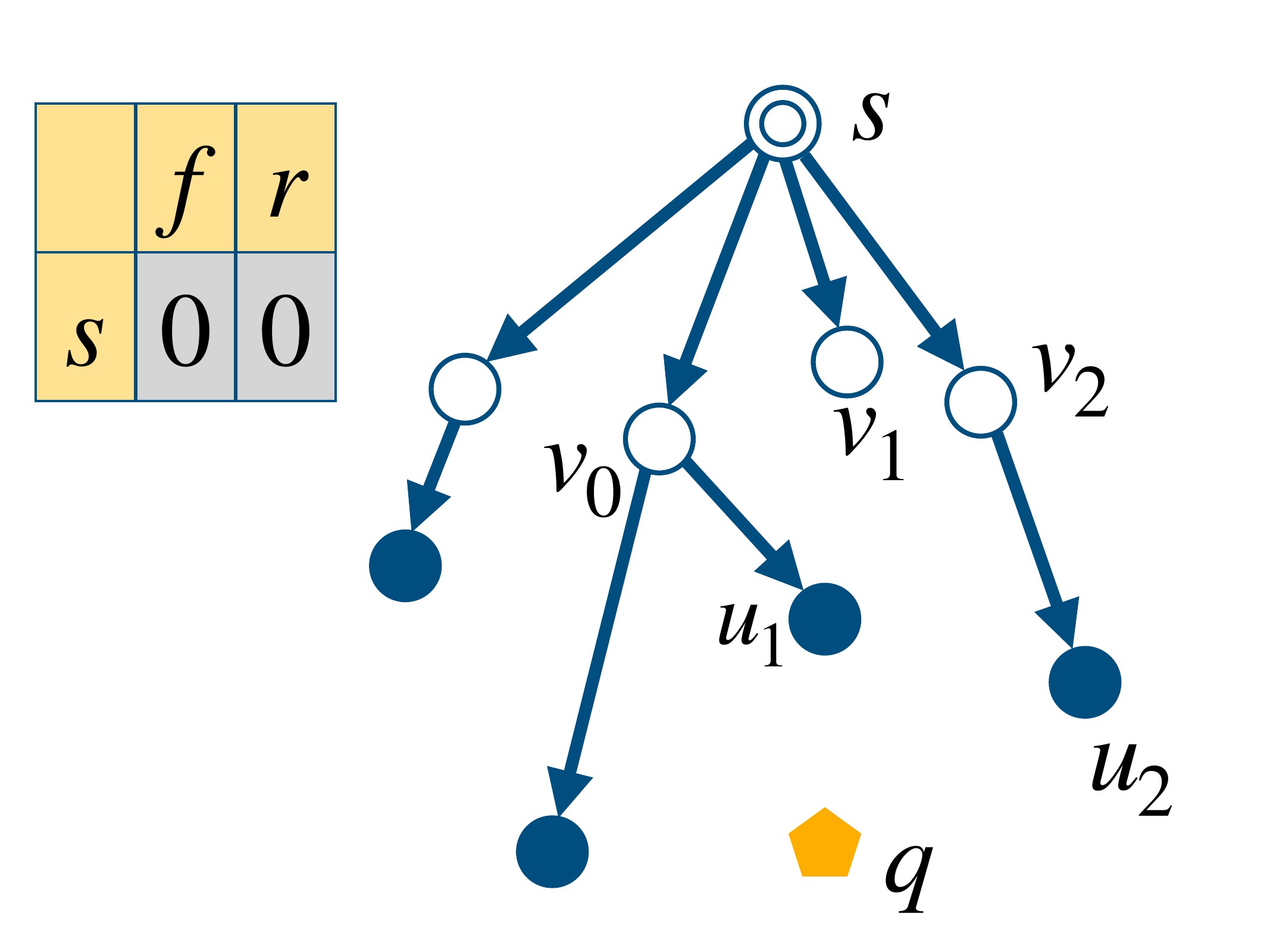}
        \caption{Step 1}
        \label{subfig:bridge_explanation_1}
    \end{subfigure}
    \begin{subfigure}[b]{0.245\linewidth}
        \centering
        \includegraphics[width=\linewidth]{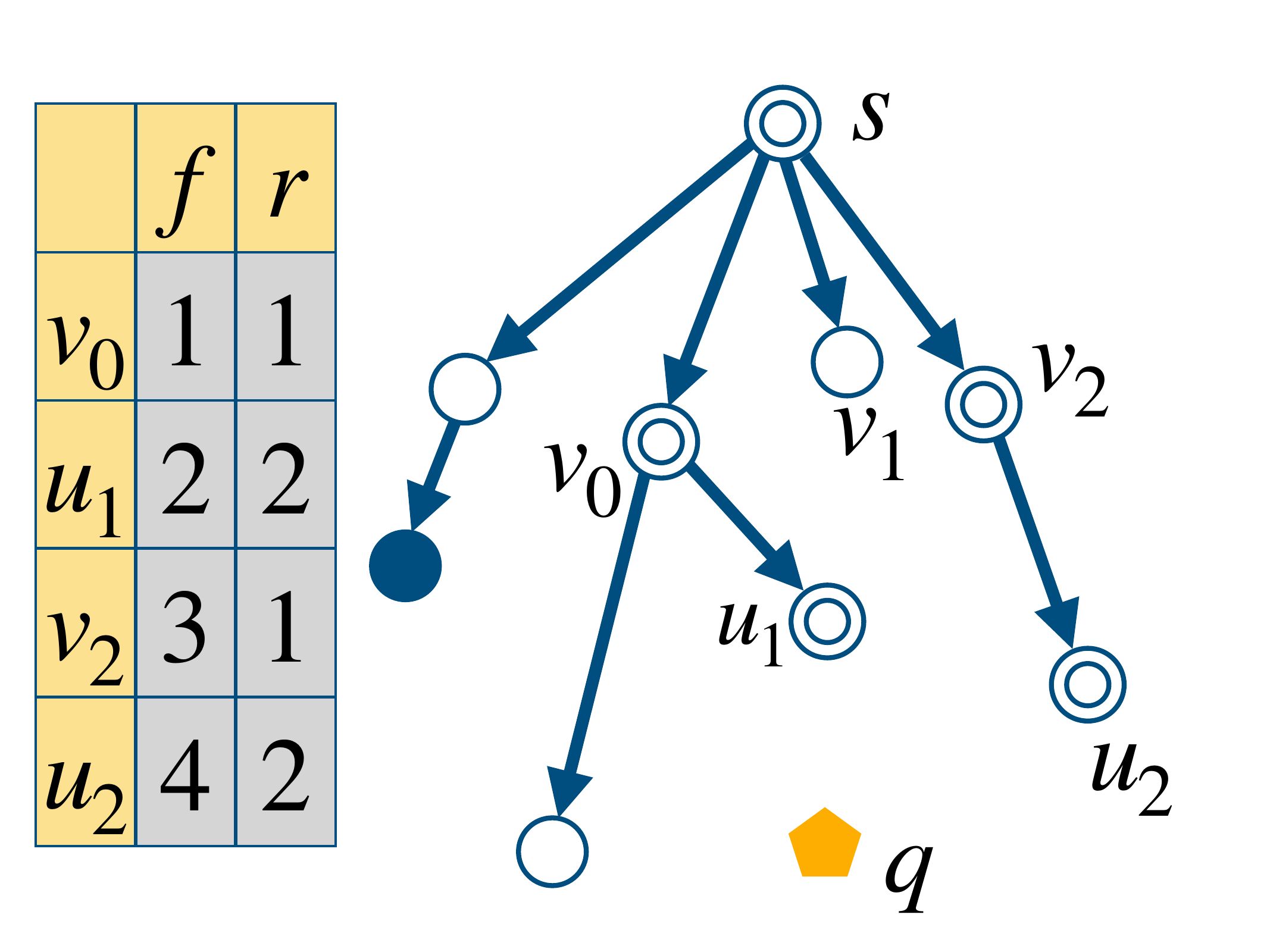}
        \caption{Step 2}
        \label{subfig:bridge_explanation_2}
    \end{subfigure}
    \begin{subfigure}[b]{0.245\linewidth}
        \centering
        \includegraphics[width=\linewidth]{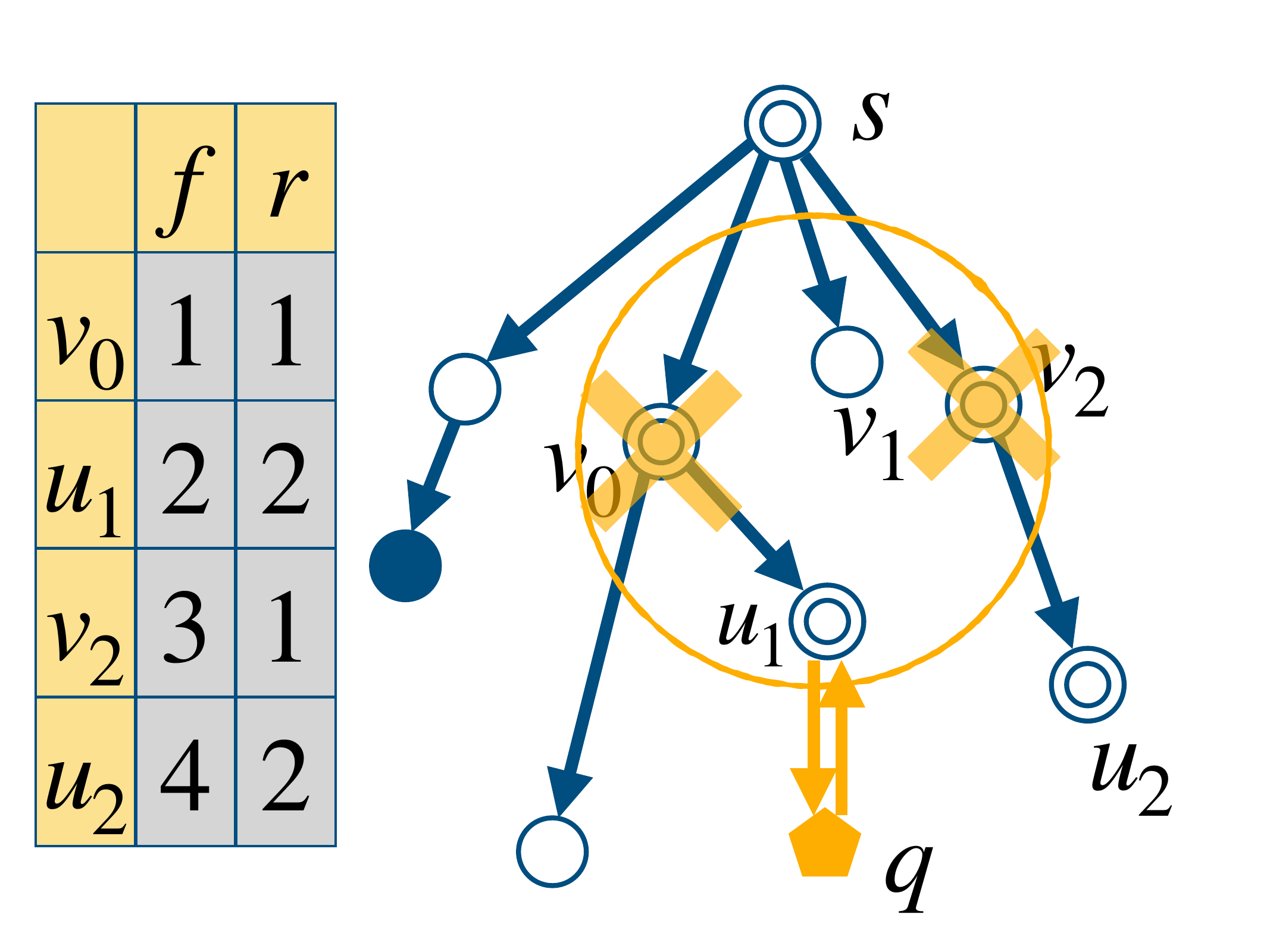}
        \caption{Step 3}
        \label{subfig:bridge_explanation_3}
    \end{subfigure}
    \begin{subfigure}[b]{0.245\linewidth}
        \centering
        \includegraphics[width=\linewidth]{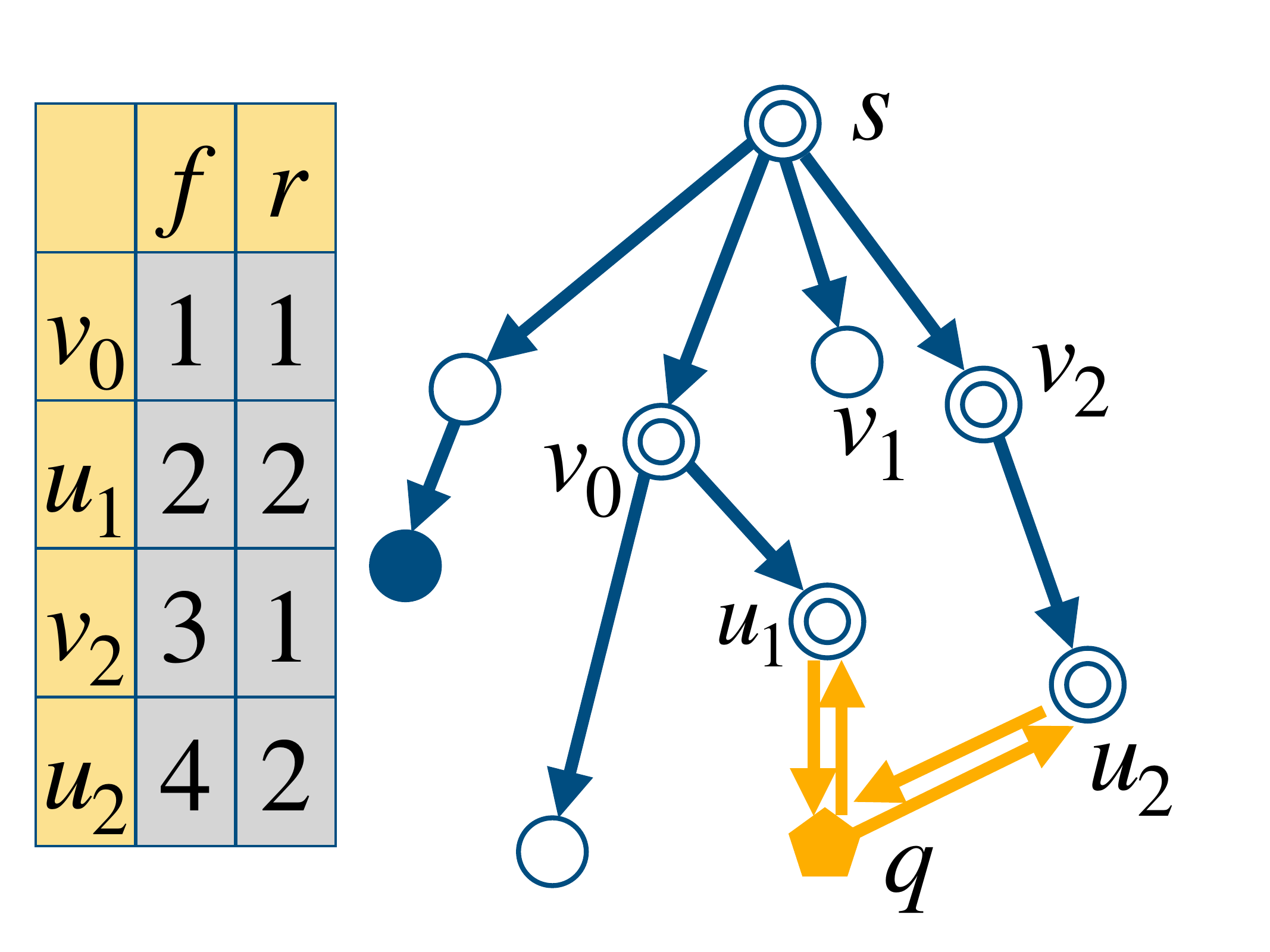}
        \caption{Step 4}
        \label{subfig:bridge_explanation_4}
    \end{subfigure}
    \begin{subfigure}[b]{0.245\linewidth}
        \centering
        \includegraphics[width=\linewidth]{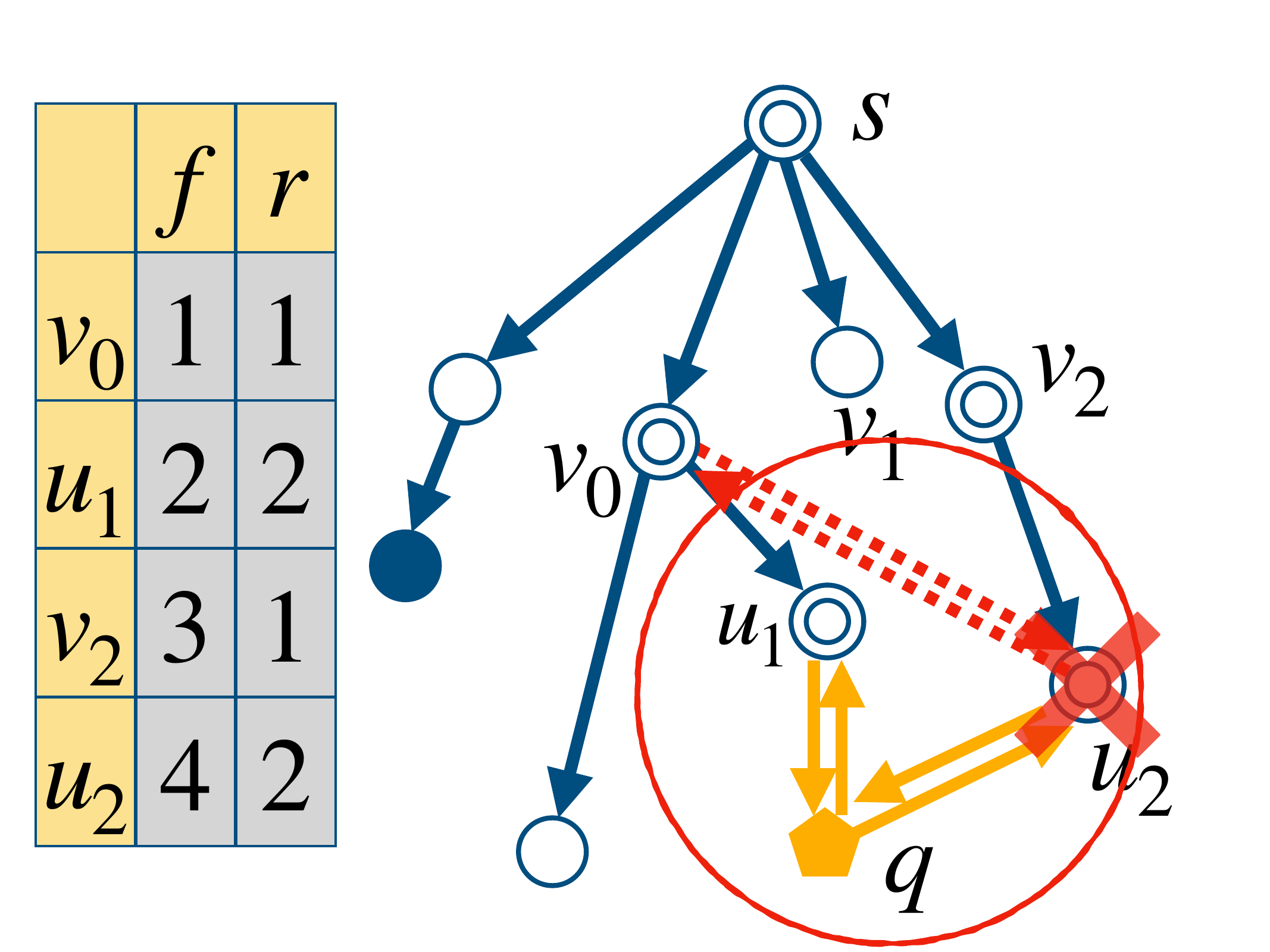}
        \caption{Step 5}
        \label{subfig:bridge_explanation_5}
    \end{subfigure}
    \begin{subfigure}[b]{0.245\linewidth}
        \centering
        \includegraphics[width=\linewidth]{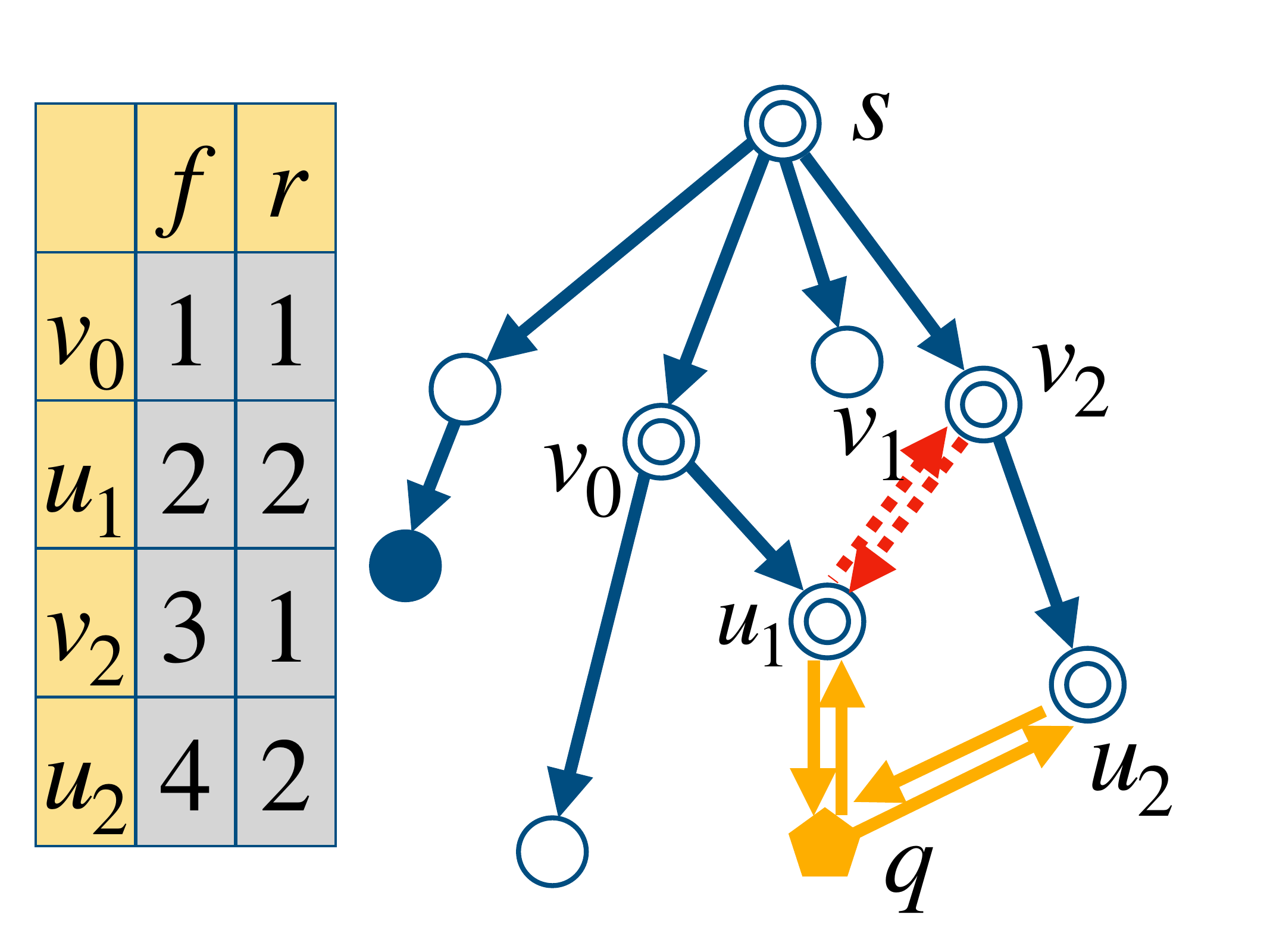}
        \caption{Step 6}
        \label{subfig:bridge_explanation_6}
    \end{subfigure}
    \begin{subfigure}[b]{0.245\linewidth}
        \centering
        \includegraphics[width=\linewidth]{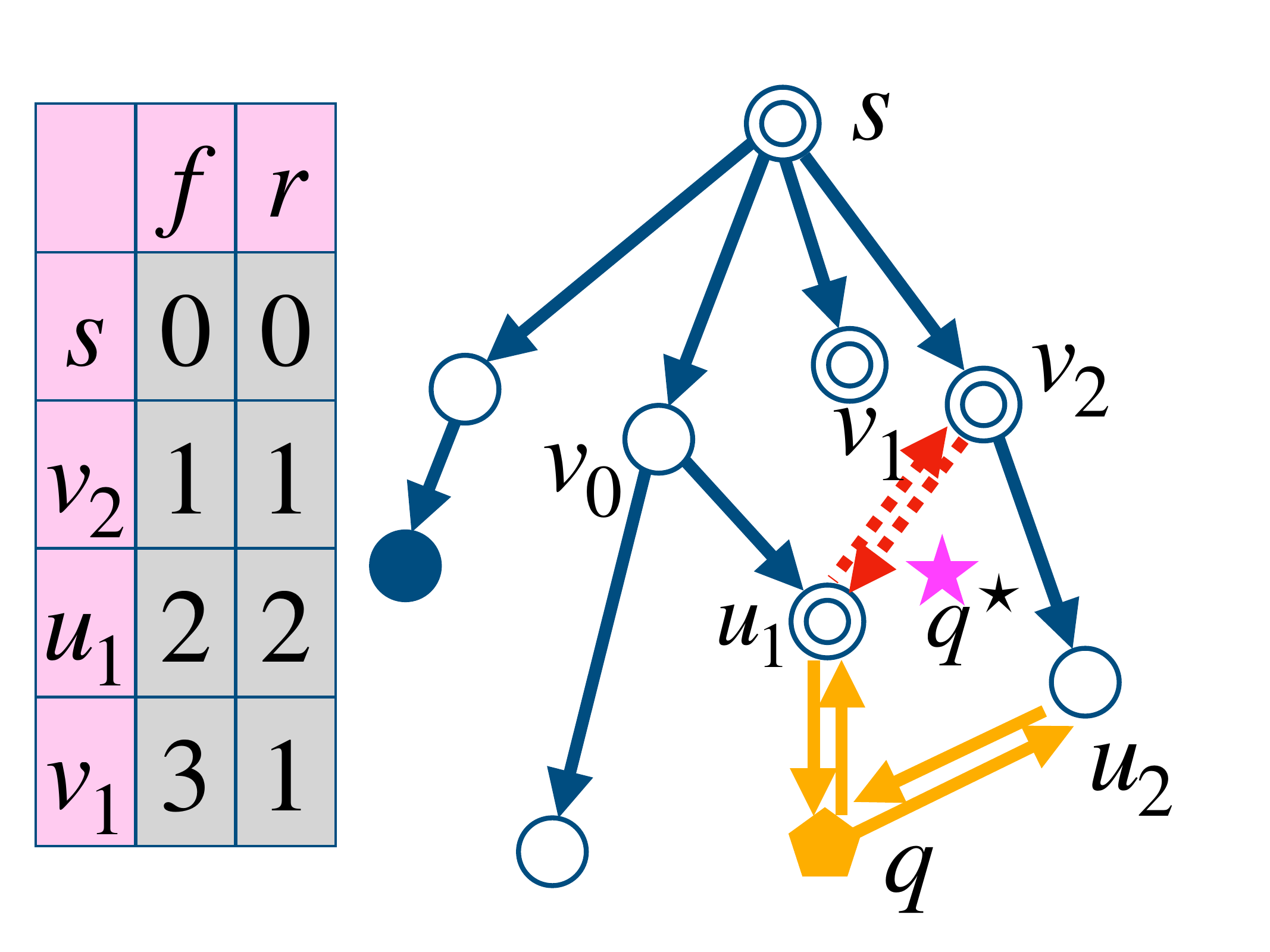}
        \caption{Step 7}
        \label{subfig:bridge_explanation_7}
    \end{subfigure}
    \begin{subfigure}[b]{0.245\linewidth}
        \centering
        \includegraphics[width=\linewidth]{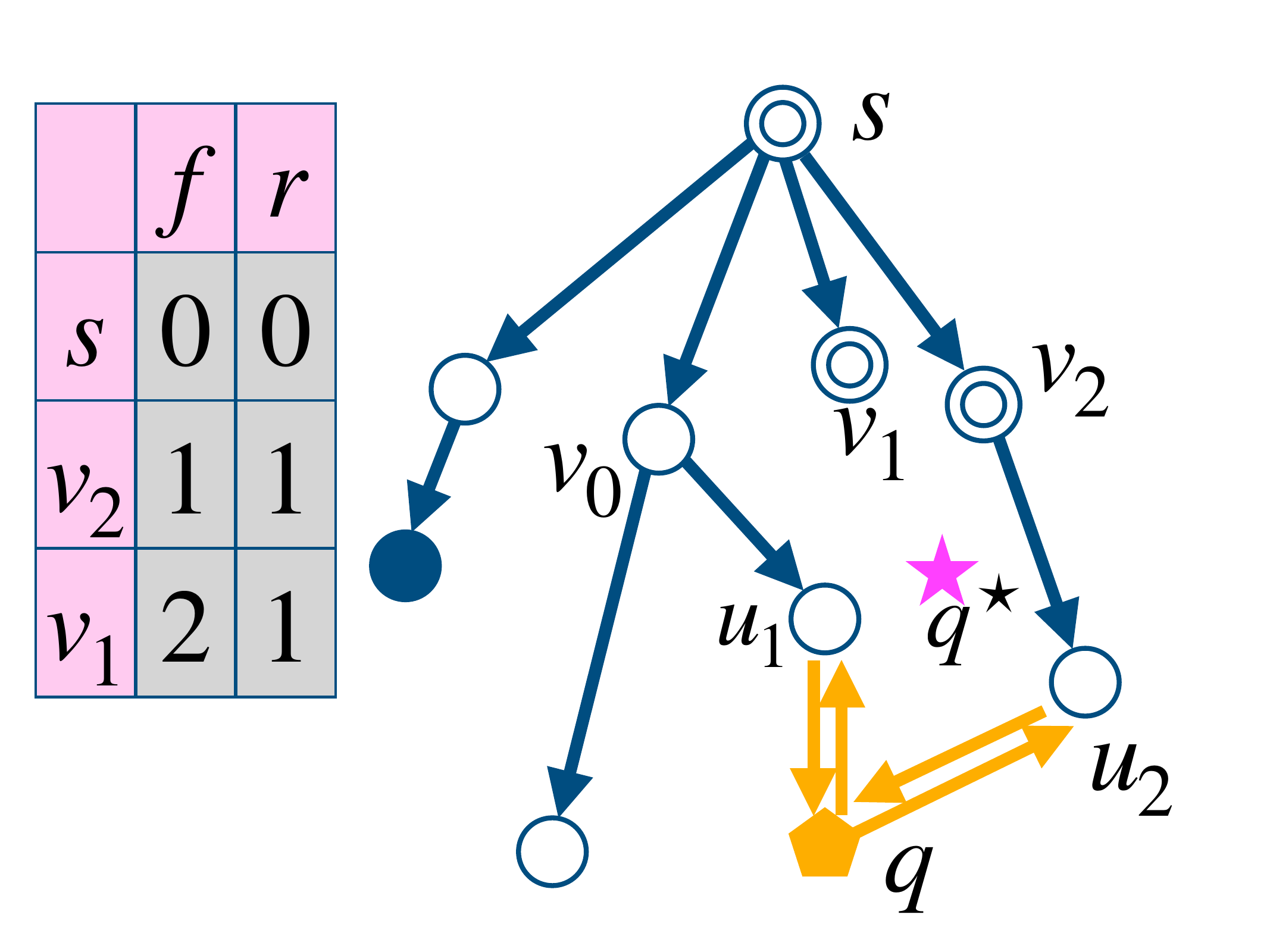}
        \caption{Step 8}
        \label{subfig:bridge_explanation_8}
    \end{subfigure}
    \caption{Step-by-step illustration of \algoname{RobustInsert} (\Cref{algorithm:robust_insert}). Solid blue edges are existing graph edges. Single hollow circles are the search frontier and double hollow circles are explored nodes. Solid yellow edges are graph edges added from inserting $q$, and dotted edges are bridge edges from inserting $q$. The $\times$ marks in Subfigures~\ref{subfig:bridge_explanation_3} and~\ref{subfig:bridge_explanation_5} are neighbor candidates pruned out by \algoname{RobustPrune}.}
    \label{figure:robust_insert_running_example}
\end{figure*}

\begin{algorithm}[t]
    \caption{Guided Bridge Building}
    \label{algorithm:guided_bridge_building}
    \small
    \begin{algorithmic}[1]
        \State $\calS$: Set of tree depths considered
        \State \algoname{HeuristicPredicate}: Boolean function for imposing constraints on the bridge edges
        \State $r(v) \equiv \text{depth of }v \text{ in } \calT$
        \State $\varname{StartIds}:$ Fixed or separately computed node set where the  search starts
        \Procedure{BridgeBuilderSearch}{$q$}
            \State \varname{frontier} $= \varname{StartIds}$ \label{line:bbbs_init_start}
            \State $\calV = \emptyset$ \Comment{Explored nodes}
            \State $\calT = \emptyset$ \Comment{Search tree}

            \For{$s\in \varname{StartIds}$}
                \State Set $\pi(s)$ = \varname{null} in $\calT$
            \EndFor
            \State $\calL \equiv$ best $L$ nodes ever seen in \varname{frontier} \label{line:bbbs_init_end}
            \While{\varname{frontier} $\cap \calL \neq \emptyset$} \label{line:bbbs_while_loop}
                \State $w = \argmin_{v_x\in \varname{frontier}\cap\calL} d(x, q)$ \label{line:best_unexplored_2}
                \State $\calV = \calV\cup\{w\}$ \label{line:bbbs_explore_start}
                \State \varname{frontier} $=$ \varname{frontier} $\setminus \{w\}$ \label{line:bbbs_explore_2}
                \For{$u\in N(w), u\not\in\calV$}
                    \State $\varname{frontier} = \varname{frontier} \cup \{u\}$ \label{line:bbbs_explore_end}
                    \State Set $\pi(u) = w$ in $\calT$\label{line:bbbs_record_tree}
                \EndFor
            \EndWhile
            \State \algoname{GuidedBridgeBuild}($\calT$) \label{line:bbbs_call_gbb}
            \State \Return $\calL, \calV$ \label{line:bbbs_return}
        \EndProcedure
        \Procedure{GuidedBridgeBuild}{$\calT$}
        \For{$v\in\calT$ such that $r(v)\in \calS$}\label{line:gbb_candidate_pairs}
            \State $\calC = \emptyset$
            \For{$w\in\calT$ such that $r(w) \in \calS$, and $w\neq v$}
                \If{\algoname{HeuristicPredicate}($v, w$)}\label{line:gbb_heuristic_predicate}
                \label{line:gbb_add_bridge}
                \State $\calC = \calC\cup\{w\}$ \label{line:gbb_candidate_pairs_2}
                \EndIf
            \EndFor
        \State \algoname{AddNeighbors}($v, \calC$) (\Cref{algorithm:add_neighbors}) \label{line:gbb_add_neighbors}
        \EndFor
        \EndProcedure
    \end{algorithmic}
\end{algorithm}

\begin{algorithm}[t]
    \caption{Add Neighbors}
    \label{algorithm:add_neighbors}
    \small
    \begin{algorithmic}[1]
        \State $R$: Global out-degree bound
        \Procedure{AddNeighbors}{$v, \calC$}
            \State $\calN = N(v) \cup \calC$
            \If{$|\calN| \leq R$}  $N(v) = \calN$
            \Else \,\, $N(v) = $ \algoname{RobustPrune}($v, \calN$) (Algorithm 3,~\cite{freshdiskann})
            \EndIf
        \EndProcedure
    \end{algorithmic}
\end{algorithm}

\begin{algorithm}[t]
    \caption{Robust Insert}
    \label{algorithm:robust_insert}
    \small
    \begin{algorithmic}[1]
        \State $L_I$: Insert backtracking budget
        \Procedure{RobustInsert}{$v_x$}
            \State $\calL, \calV = \algoname{BridgeBuilderSearch}_{L=L_I}(x)$ (\Cref{algorithm:guided_bridge_building})
            \State $N(v_x) = $ \algoname{RobustPrune}($v_x, \calV$) (Algorithm 3,~\cite{freshdiskann})  \label{line:robust_insert_prunes}
            \For{$w\in N(v_x)$}\label{line:robust_insert_reverse_edges}
                 \algoname{AddNeighbors}($w, \{v_x\}$)  (\Cref{algorithm:add_neighbors})
            \EndFor
        \EndProcedure
    \end{algorithmic}
\end{algorithm}

\subsubsection{\algoname{Insert} with \algoname{GuidedBridgeBuild}}~\Cref{figure:robust_insert_running_example} illustrates \algoname{RobustInsert}, which uses \algoname{GuidedBridgeBuild} with insert queries. In the example, $L=2$ and $R=2$, and degree constraint $R$ does not apply to the starting node $s$. $r(\cdot)$ is the depth in the search tree and $f(\cdot)$ is the order in which nodes are explored. $q$ is the new data point being inserted. Steps 1--2 (Subfigures \ref{subfig:bridge_explanation_1}--\ref{subfig:bridge_explanation_2}) explore $\calV=\{s, v_0, u_1, v_2, u_2\}$. On Step 2, the search terminates with $L=2$ and the local minima of $d(\cdot, q)$: $u_1$ and $u_2$. On Line~\ref{line:robust_insert_prunes} in~\Cref{algorithm:robust_insert}, \algoname{RobustInsert} uses \algoname{RobustPrune} to select a subset of neighbors for $q$ from $\calV$. 
In Step 3 (Subfigure~\ref{subfig:bridge_explanation_3}), \algoname{RobustPrune} selects $u_1$ as the closest neighbor candidate, pruning $v_0$ and $v_2$. $u_2$ is subsequently selected as the next neighbor, reaching the out-degree bound $R=2$. $q$ is therefore connected to $u_1$ and $u_2$ (Subfigure ~\ref{subfig:bridge_explanation_4}).
Next, \algoname{GuidedBridgeBuild} with $\calS=\{1, 2\}$ and no \algoname{HeuristicPredicate} 
considers bridge edges between nodes satisfying $r(\cdot)\in\calS$: edges $v_0\leftrightarrow u_2$ and $u_1\leftrightarrow v_2$. Subfigure ~\ref{subfig:bridge_explanation_5} illustrates $u_1$ causing $v_0\rightarrow u_2$ to be pruned. Similarly, $v_2$ causes $u_2\rightarrow v_0$ to be pruned. The edges $u_1\leftrightarrow v_2$ are added as bridge edges (Subfigure ~\ref{subfig:bridge_explanation_6}), concluding \algoname{RobustInsert}.

This example also shows the algorithmic benefits of \algoname{GuidedBridgeBuild}. Subfigure~\ref{subfig:bridge_explanation_7} illustrates a search for $q^\star$ with the bridge edges present in the graph. The nodes $s$, $v_2$, $u_1$, and $v_1$ are explored in this order, correctly finding $u_1$, the correct nearest neighbor for $q^\star$. Subfigure~\ref{subfig:bridge_explanation_8} illustrates the search for $q^\star$ without bridge edges. $s$, $v_2$, and $v_1$ are explored, but the search terminates at the incorrect local optima, $v_1$ and $v_2$. $v_0$, which navigates to $u_1$, is too far from the query compared to the other siblings of $v_0$. With the same backtracking budget as when the search uses bridge edges, $v_0$ and $u_1$ are missed in the search. Therefore, the bridge edges between $u_1$ and $v_2$ 
improve the navigability of the graph.

\Cref{algorithm:robust_insert} lists our insertion algorithm \algoname{RobustInsert}. Given a new node $v_x$, \algoname{RobustInsert} first runs \algoname{BridgeBuilderSearch} for $x$, and then selects $N(v_x)$ from $\calV$ (nodes explored by \algoname{BridgeBuilderSearch}) and tries to add $v_x$ to $N(w)$ for each $w\in N(v_x)$, using \algoname{RobustPrune} as needed.

\subsubsection{\algoname{Search} with \algoname{GuidedBridgeBuild}} A \algoname{Search} query can also improve the part of the index that it traverses with \algoname{BridgeBuilderSearch}. Searches for other nearby points traverse similar parts of the graph and thus benefit from the improved navigability. We refer to \algoname{Search} queries that perform \algoname{GuidedBridgeBuild} as \textit{training} queries hereafter. In \Cref{section:experiments}, we find experimentally that training with even a small number of in-distribution queries helps the index adapt to distribution shifts in a query-aware fashion.

\subsubsection{Hyperparameters $\calS$ and \algoname{HeuristicPredicate}} \label{subsubsec:bridge_hyperparams}
We found experimentally that $\Theta(\log(|{D}|))$ with a $\Theta(1)$ width is a good range for $\calS$. \algoname{HeuristicPredicate} further filters out candidate bridge edges to balance between the cost of potentially performing more \algoname{RobustPrune}s for bridge edges and the robustness benefits these additional connections provide.  
By default, \algoname{HeuristicPredicate} in our experiments requires two endpoints of a bridge to have the same depth in $\calT$. We also compare this heuristic to several alternatives in \Cref{section:experiments}.

\subsubsection{Summary} Our bridge building algorithm has three key benefits: (1) indexes whose construction and insertions use our algorithm have increased search quality under the same backtracking budget compared to \sysname{FreshVamana}, achieving a better search quality-efficiency tradeoff; (2) in a sliding window update scenario, our algorithm consistently improves the recall of the index throughout updates; and (3) our method can effectively adapt to distribution shifts and out-of-distribution queries. One can adapt \algoname{GuidedBridgeBuild} to other graph-based indexes by plugging in their backtracking and neighbor pruning mechanisms.

\section{Dynamic Concurrent Data Structure Cleaning}\label{section:dynamic_consolidate}
In this section, we discuss how to handle \algoname{Delete}s concurrently with \algoname{Insert}s and \algoname{Search}es. We guarantee that a client will not see a deleted point in the results of a \algoname{Search} query executed after the \algoname{Delete} has completed. 
To achieve this consistency guarantee, it suffices to mark a deleted point as a \textbf{\emph{tombstone}}, run subsequent \algoname{GreedySearch}es normally, and filter out the tombstones when selecting the $k$ best neighbors from $\calL$. However, as the index becomes more bloated, the accumulated tombstones interfere with the graph traversal, leading to a degradation in recall, as observed by~\citet{spfresh} and in our experiments. Therefore, to achieve high search quality, we need to remove the tombstones and repair the graph. Previous graph-based ANNS indexes use costly global operations to implement graph repairs~\cite{freshdiskann, hezel2023_deg, lsh_apg}. If the graph repair or rebuild operations are performed in batches, more expensive searches with larger backtracking budgets (e.g., the \sysname{pgvector HNSW} iterative scan~\cite{pgvector_iterative_scan}) are required to maintain recall. To address these issues, we propose adaptive dynamic techniques that perform graph repairs and cleaning more efficiently.

\begin{algorithm}[t]

    \caption{Consolidation (Subroutine of Algorithm 2 in~\cite{freshdiskann})}
    \small
    \label{algorithm:consolidation}
    \begin{algorithmic}[1]
        \State $R:$ Global out-degree bound
        \Procedure{Consolidate}{$v$}
        \State $\calC = \emptyset$ \Comment{Neighbor candidates}
        \For{$w \in N(v)$}
            \If{\algoname{IsLive}($w$)}
                \ $\calC = \calC \cup \{w\}$
            \Else \ 
                 $\calC = \calC \cup \{u\in N(w)\mid u\neq v \land \algoname{IsLive}(u)\}$
            \EndIf
        \EndFor
        \If {$|\calC| \leq R$}
            $N(v) = \calC$
        \Else \
            $N(v) = \algoname{RobustPrune}(v, \calC)$
        \EndIf
        
        \EndProcedure
    \end{algorithmic}
\end{algorithm}


\subsection{Algorithm Description}
\label{subsec:consolidate_algo_descriptions}

We first outline the life cycle of any graph node and the corresponding states:

\begin{itemize}[leftmargin=*]
    \item \textbf{Live}: The node represents a valid data point.
    \item \textbf{Tombstone}: The data corresponding to the node is logically deleted and will not be returned by a \algoname{Search} query. The graph in- and out-neighborhoods and the data value of the node are still valid for graph traversal.
    \item \textbf{Replaceable}: The node slot is available in $M$ for a new insertion. The corresponding data and the out-neighborhood are no longer valid but there may be remaining in-neighbors.
\end{itemize}

Throughout the discussion, let $w_x$ be the tombstone node to be removed.
Concretely,  \algoname{Delete} had previously been called on $x$, the index has marked $w_x$ as a tombstone, and the graph needs to be repaired to prepare for the actual deletion of $w_x$.
We propose performing neighborhood consolidation on the fly when needed and an accompanying semi-lazy memory management strategy.
The discussion and implementation are based on \sysname{Vamana}, but the techniques are applicable to any graph-based ANNS index.

\myparagraph{Efficient Consolidation on the Fly}  \algoname{GreedySearch} can detect tombstone nodes used for its navigation. If \algoname{GreedySearch} adds $w_x$ to the search frontier when exploring $v$, i.e., $v = \pi(w_x)$ in $\calT$, and $v$ is live, then the edge $v\rightarrow w_x$ facilitated the $v\rightarrow w_x\rightarrow N(w_x)$ navigation. Therefore, it is useful to make $v$ absorb $N(w_x)$ via neighborhood consolidation before we actually delete $w_x$. This can be carried out by either \algoname{Insert} or \algoname{Search} queries. Since all of $v$'s unvisited tombstone neighbors are added to the frontier during the same search iteration, it is cost-effective for $v$ to consolidate with all of its tombstone neighbors' out-neighborhoods (\algoname{Consolidate} in \Cref{algorithm:consolidation}). This makes tombstone nodes ready to be removed from the graph earlier.

\myparagraph{Early Stopping of Consolidations} Immediately removing $w_x$ after only one in-neighbor $v$ detects and consolidates with it does not sufficiently repair the graph around $w_x$. However, waiting for all incoming neighbors of $w_x$ to reach $w_x$ through a \algoname{GreedySearch} and consolidate with $w_x$ reduces to naively amortizing \sysname{FreshVamana}'s expensive periodic consolidation of connecting all in-neighbors to all out-neighbors of a deleted node.
Since our method chooses effective consolidation targets on the fly, we find that it is unnecessary to connect all nodes in $IN(w_x)$ to all nodes in $N(w_x)$, and
that it suffices to visit a tombstone $w_x$ just a few times with \algoname{Consolidate}($v$) ($v=\pi(w_x)$ in some search tree $\calT$) to preserve $w_x$'s navigation functionality. Therefore, we maintain a count for each tombstone of the number of times it has been visited by \algoname{Consolidate} from a live parent and stop performing more \algoname{Consolidate}s for $w_x$ after the count hits a threshold $C$ (which we call the eagerness threshold). We present an analysis of the sensitivity of $C$ in~\Cref{section:experiments}.

\myparagraph{Semi-Lazy Cleaning}
The remaining issue is to remove a tombstone node $w_x$ from the graph after it has been processed by live parent nodes a sufficient number of times with \algoname{Consolidate}.
Existing methods (\sysname{FreshVamana}~\cite{freshdiskann}, \sysname{IP-DiskANN}~\cite{xu2025inplaceupdatesgraphindex}, and \sysname{pgvector HNSW}~\cite{pgvector}) remove dangling edges $(w_i\rightarrow w_x)$ for any remaining $w_i\in IN(w_x)$ that has not consolidated with $w_x$ with costly batch consolidations.
We argue that cleaning the remaining dangling incoming edges is \textit{not necessary}. The primary benefits of actually removing $w_x$ and all of its \textit{incoming} edges include
(1) the storage for node $w_x$ can be reused for future nodes that are inserted and 
(2) $w_x$'s in-neighbors that did not call \algoname{Consolidate} on $w_x$'s neighborhood
do not unnecessarily explore $w_x$ in future \algoname{GreedySearch}es, avoiding potential incorrect local optima. However, if previous on-the-fly \algoname{Consolidate}s for $w_x$ already sufficiently repaired the graph around $w_x$, it is acceptable for a deleted point $x$ to be \textit{directly overwritten by a new data point $y$}, even when some in-neighbors still point to the old $w_x$ node in the graph.

When a new data point $y$ is inserted, if $w_x$ has been \algoname{Consolidate}d $C$ times, then \textit{\algoname{RobustInsertData} can turn $w_x$ into $w_y$, representing $y$ directly}. $\algoname{RobustInsert}(w_y)$ then uses $y$ in its distance computations and neighbor selection. The resulting $IN(w_y)$ contains both new in-neighbors from $\algoname{RobustInsert}(w_y)$ and the remaining nodes from $IN(w_x)$. We call the remaining incoming edges from $IN(w_x)$ \textbf{\emph{random edges}}, with the randomness coming from the workload. 
We experimentally confirm 
that these random edges do not hurt the quality of the index, since \algoname{GreedySearch} naturally avoids exploring nodes that are far away, and \algoname{RobustPrune} naturally removes edges that are not useful. We call this technique \textbf{\textit{semi-lazy cleaning}}.


\begin{algorithm}[t]
    \caption{Try Marking a Node as Replaceable}
    \label{algorithm:try_mark_replaceable}
    \small
    \begin{algorithmic}[1]
        \State $M$: Linearizable bag data structure tracking available graph storage slots corresponding to replaceable nodes
        \State $C$: Eagerness threshold
        \State $H$: Per-node atomics for tracking node status \label{line:cdbs_h_init}
        \Comment{For any node $w$, $H(w) \geq 0$ iff $w$ is a \textbf{tombstone} and $w$ is not \textbf{replaceable}; $H(w) = -2$ iff $w$ is \textbf{replaceable}; $H(w) = -1$ iff $w$ is \textbf{live}} \label{line:cdbs_init_tombstone_dict}
        
        \Procedure{TryMarkReplaceable}{$w$}
            \While{\textbf{true}}
                \State \varname{current} $\leftarrow$ \algoname{Load}($H(w)$)
                \If{\varname{current} $< C$}
                    \Return \varname{current} $== -2$ 
                \EndIf
                \If{CAS($H(w)$, \varname{current}, $-2$)}
                    \State $M\leftarrow M\cup \{w\}$
                    \State \Return 
                    \textbf{true}
                \EndIf
            \EndWhile
        \EndProcedure
    \end{algorithmic}
    
\end{algorithm}

\begin{algorithm}[t]
    \caption{Clean Dynamic Search}
    \label{algorithm:clean_dynamic_beam_search}
    \small
    \begin{algorithmic}[1]
        \State $L$: Search backtracking budget
        \State $\varname{StartIds}:$ Fixed or separately computed node set where the  search starts
        \State $M$: Linearizable bag data structure tracking available graph storage slots corresponding to replaceable nodes
        \State $C$: Eagerness threshold
        \State $H$: Per-node atomics for tracking node status \label{line:cdbs_h_init}
        
        \Procedure{CleanDynamicSearch}{$q$, $\varname{performance\_sensitive}$}
            \State $\calT$: Data structure for maintaining the search tree \label{line:cdbs_init_start}
            \State \varname{frontier} $= \varname{StartIds}$
            \State $\calL = \varname{StartIds}$ \Comment{Maintains the best $L$ nodes visited}
            \State $\calV = \emptyset$ \Comment{Explored nodes}
             \label{line:cdbs_init_end}
            
            \While{\varname{frontier} $\cap \calL \neq \emptyset$}
                \State $w = \argmin_{v_x\in \varname{frontier}\cap\calL} d(x, q)$ \label{line:cdbs_explore_1}
                \State $\calV = \calV\cup\{w\}$ 
                \State Remove $w$ from \varname{frontier}\label{line:cdbs_explore_2}
                \State $\calN = \{u\in N(w) \mid u\not\in\calV, \neg\algoname{TryMarkReplaceable}(u)\}$ \Comment{Calls \algoname{TryMarkReplaceable}($u$) on each $u$ in $N(w)$ to evaluate whether $u$ is replaceable, and keeps nodes with $H(u)>-2$}\label{line:cdbs_free}
                \For{$u \in \calN$} \label{line:cdbs_explore_4}
                    \State $\varname{frontier} = \varname{frontier}\cup\{u\}$

                    \State $\calL = \calL\cup\{u\}$
                    \State Delete $\argmax_{v_x\in\calL}d(q, x)$ from $\calL$ if $|\calL| > L$
                \EndFor

                \For{$u\in\calN$}
                \State Set $w = \pi(u)$ in $\calT$\label{line:cdbs_record_tree}
                \EndFor
                
                \If {$\exists u\in \calN,\ \algoname{IsTombStoned}(u) \land \neg \algoname{IsTombStoned}(w)$} \label{line:cdbs_consolidate}
                    \State $\algoname{CleanConsolidate}(w)$ (\Cref{algorithm:clean_consolidation}) \label{line:cdbs_consolidate2}
                \EndIf

            \EndWhile
            \If{$\neg \varname{performance\_sensitive}$}
            \State \algoname{GuidedBridgeBuild}($\calT$) (\Cref{algorithm:guided_bridge_building}) \label{line:cdbs_gbb}
            \EndIf
            \State \Return $\calL, \calV$
        \EndProcedure
    \end{algorithmic}
\end{algorithm}

\begin{algorithm}[t]
    \caption{Clean Consolidation }
    \label{algorithm:clean_consolidation}
    \small
    \begin{algorithmic}[1]
        \State $H$: Per-node atomics for tracking node status 
        \State $w$: Tombstone node whose consolidation counter is updated
        \State $r$: Number of new consolidations
        \Procedure{RecordConsolidation}{$w$, $r$}
            \While{\textbf{true}}
                \State \varname{current} $\leftarrow$ \algoname{Load}($H(w)$)
                \If{\varname{current} $< 0$}
                    \Return
                \EndIf
                \If{CAS($H(w)$, \varname{current}, \varname{current + $r$})}
                    \Return
                \EndIf
            \EndWhile
        \EndProcedure
        \Procedure{CleanConsolidate}{$v$}
        \State $\calN = \{w\in N(v) \mid \algoname{IsTombStoned}(w)\}$
        \State $\algoname{Consolidate}(v)$\label{line:cc_consolidate}
        \For{$w\in \calN$}\label{line:cc_increment_h}
            \ \algoname{RecordConsolidation}($w$, 1)
        \EndFor
        \EndProcedure
    \end{algorithmic}
\end{algorithm}
\begin{algorithm}[t]
    \caption{\system{} Insert Data}
    \label{algorithm:cleann_insert_data}
    \small
    \begin{algorithmic}[1]
        \State $H$: Per-node atomics for tracking node status
        \State $M$: Linearizable bag data structure tracking available graph storage slots corresponding to replaceable nodes
        \Procedure{RobustInsertData}{$x$}
            \State $w\leftarrow$ Pop any element in $M$ 
            \State \varname{current} $\leftarrow$ \algoname{Load}$(H(w))$
            \State assert($\varname{current} == -2$)
            \State $H(w)\leftarrow -1$
        \State \algoname{RobustInsert}($w_x$) (\Cref{algorithm:robust_insert})
        \EndProcedure
    \end{algorithmic}
\end{algorithm}
\begin{algorithm}[t]
    \caption{\system{} Search}
    \label{algorithm:cleann_search}
    \small
    \begin{algorithmic}[1]
        \State $H$: Per-node atomics for tracking node status
        \State $\varname{performance\_sensitive}$: Boolean variable indicating whether the search is performance-sensitive.
        \Procedure{Search}{$k$, $q$, $\varname{performance\_sensitive}$}
        \State $\calL, \_$ = \algoname{CleanDynamicSearch}($q, \varname{performance\_sensitive}$)
        \State $\calL' = \{v\in \calL\mid H(v) = -1\}$\label{line:cs_filter_tombstone}
        \State \Return $k$ best points corresponding to nodes in $\calL'$ using brute force comparison with $q$
        \EndProcedure
    \end{algorithmic}
\end{algorithm}
\begin{algorithm}[t]
    \caption{\system{} Delete}
    \label{algorithm:cleann_delete}
    \small
    \begin{algorithmic}[1]
        \State $H$: Per-node atomics for tracking node status 
        \Procedure{Delete}{$w$} 
        \While{\textbf{true}}
            \State \varname{current} $\leftarrow$ \algoname{Load($H(w)$)}
            \If{\varname{current} $\neq -1$}
                \Return \textbf{false}
            \EndIf
            \If{CAS($H(w)$, \varname{current}, $0$)}
                \Return \textbf{true}
            \EndIf
        \EndWhile
        \EndProcedure
    \end{algorithmic}
\end{algorithm}

\myparagraph{\textbf{Algorithm}}
\Cref{algorithm:clean_dynamic_beam_search} lists \algoname{CleanDynamicSearch}, the full algorithm with on-the-fly consolidation and semi-lazy memory cleaning, combined with the methods proposed in ~\Cref{section:bridge}. Apart from the bookkeeping for the best-first search and \algoname{GuidedBridgeBuild}, we now also need the array $H$, which encodes the tombstone and consolidation count information (Line~\ref{line:cdbs_h_init}). For all live nodes $w$, we have $H(w) = -1$, and for all replaceable nodes, we have $H(w)=-2$. 
For tombstoned nodes $w$, $H(w)$ tracks the number of consolidations that have been performed for them, and is non-negative.
The code to modify $H$ uses \emph{compare-and-swap (CAS)}, which takes three arguments---a memory location $x$, an expected old value \emph{old}, and a desired new value \emph{new}.
It atomically does the following: reads the value of $x$, if it is equal to \emph{old}, then writes \emph{new} to $x$ and returns \emph{true}; otherwise, leaves $x$ unchanged and returns \emph{false}. 

During the best-first search, after the current best node $w$ is selected to be explored and marked as explored (Lines~\ref{line:cdbs_explore_1}--\ref{line:cdbs_explore_2}), for each neighbor $u$ of $w$, if $u$ is a tombstone and has been consolidated for at least $C$ times, $u$ is marked replaceable and is available to represent a new point (Line~\ref{line:cdbs_free}) via setting $H(u) = -2$. On Lines~\ref{line:cdbs_free}--\ref{line:cdbs_record_tree}, $w$ is explored, adding its unvisited out-neighbors to the frontier and recording them as its children in the search tree $\calT$. If any of these children are tombstoned nodes, the algorithm performs $\algoname{CleanConsolidate}(w)$ (\Cref{algorithm:clean_consolidation}), consolidating the neighborhoods of all of $w$'s deleted out-neighbors with $w$'s neighborhood, incrementing $H$ for the out-neighbors involved to record the consolidation, and pruning as appropriate (Lines~\ref{line:cdbs_consolidate}--\ref{line:cdbs_consolidate2}).

For queries that are performance-sensitive (i.e., not training queries), \algoname{GuidedBridgeBuild} is not called. Performance-sensitive queries still perform on-the-fly consolidation.

\begin{figure}
\vspace{-3pt}
\centering
\begin{minipage}[t]{.48\linewidth}
\includegraphics[width=\linewidth]{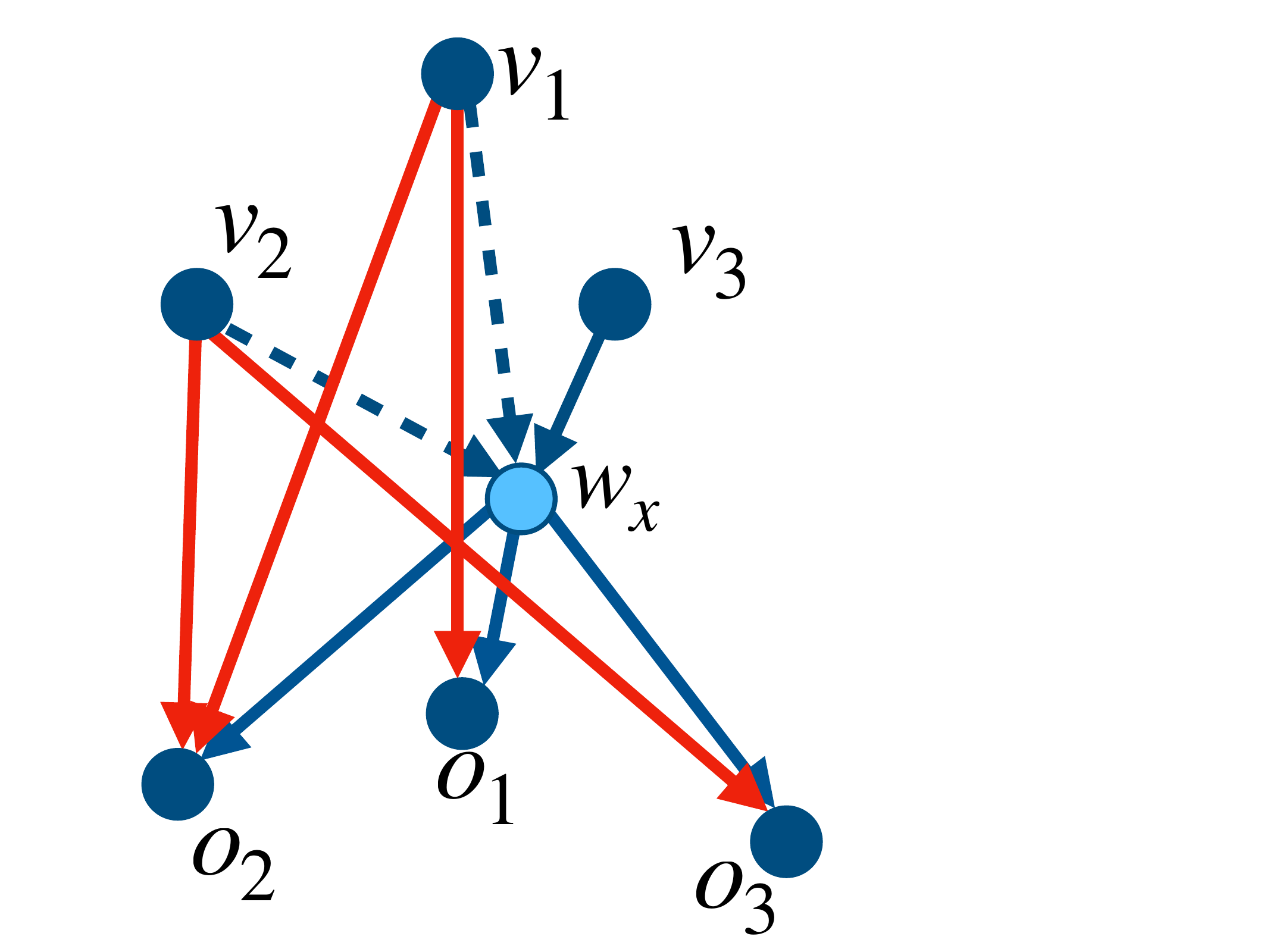} 
\caption{$x$ is the data point being deleted. Blue edges existed before \algoname{Delete}($x$). Red edges are added by \algoname{CleanConsolidate} (\Cref{algorithm:clean_consolidation}). Solid edges exist and dashed edges are deleted. $w_x$ is consolidated $C=2$ times and marked replaceable.
}\label{figure:consolidate_explanation_1}
\end{minipage}
\hfill
\begin{minipage}[t]{.48\linewidth}
\includegraphics[width=\linewidth]{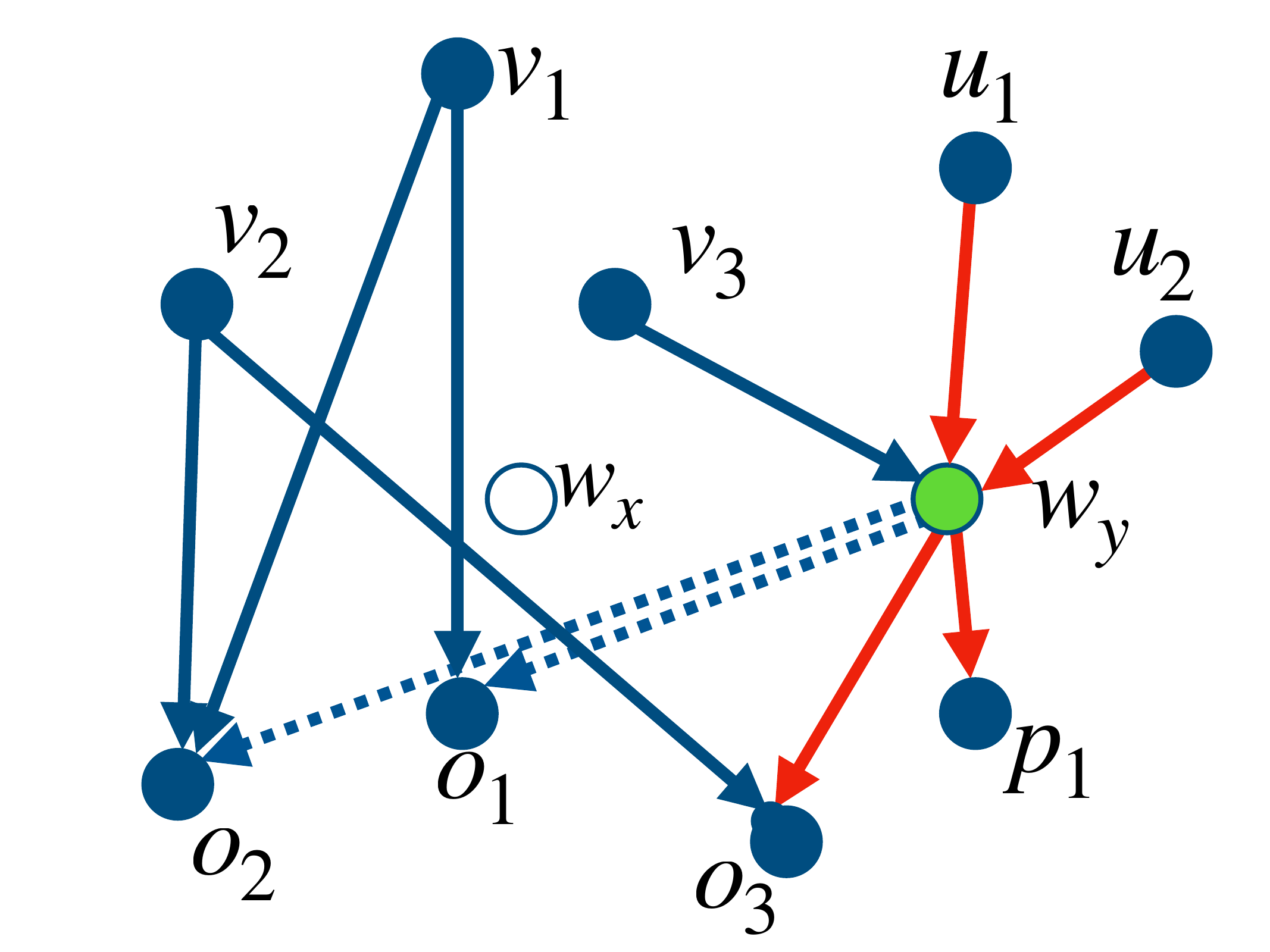} 
\caption{$y$ replaces $x$ to be the data point that $w$ represents. Blue edges existed before the replacement. Red edges are determined using $y$. Solid edges exist after $y$ is inserted and dashed edges are deleted as $w_y$'s out-neighborhood is pruned using $y$.}\label{figure:consolidate_explanation_2}
\end{minipage}
\label{figure:dynamic_consolidate_explanation}
\Description[Illustration for adaptive on-the-fly consolidation algorithm and semi-lazy memory cleaning.]{This figure is an example to help understand the algorithms described in \Cref{subsec:consolidate_algo_descriptions}.}
\end{figure}

Figures~\ref{figure:consolidate_explanation_1}--\ref{figure:consolidate_explanation_2} illustrate an example of \algoname{CleanDynamicSearch} with $C=2$.
In~\Cref{figure:consolidate_explanation_1}, initially, $N(w_x) = \{o_1, o_2, o_3\}$ and $IN(w_x) = \{v_1, v_2, v_3\}$. \algoname{Delete}($x$) marks $w_x$ as a tombstone. As queries explore $v_1$ and $v_2$ via \algoname{CleanDynamicSearch} (\Cref{algorithm:clean_dynamic_beam_search}), $v_1$ and $v_2$ consolidate with $w_x$, absorbing $\{o_1, o_2, o_3\}$ into $N(v_1)$ and $N(v_2)$. The edges $(v_1\rightarrow w_x)$ and $(v_2\rightarrow w_x)$ are deleted. Since $C=2$, $v_3$ does not perform \algoname{CleanConsolidate} anymore and the edge $v_3\rightarrow w_x$ remains. $w$ becomes available for representing other data points (it is replaceable).

In~\Cref{figure:consolidate_explanation_2}, after $y$ is inserted, $w_x$ becomes $w_y$.  $N(w_y)$ is formed with \algoname{RobustInsert} using $y$ for distance computations. During \algoname{RobustPrune} for $w_y$, both the old out-neighborhood $N(w_x)$ and the visited node set $\calV$ generated by $\algoname{CleanDynamicSearch}(y)$ are candidates for $N(w_y)$. The random edge $v_3\rightarrow w_y$ stays in the graph.

\subsection{Implementing Operations using \algoname{CleanDynamicSearch}} \Cref{algorithm:cleann_search} lists the \algoname{Search} implementation in \system{}. Similar to other graph-based indexes, it suffices to return the $k$ best points corresponding to live nodes in $\calL$ generated by \algoname{CleanDynamicSearch}. For \algoname{Delete}, it suffices to mark the node deleted by changing its $H$ value from $-1$ to $0$ (\Cref{algorithm:cleann_delete}).

\subsection{Concurrency in \system}


Under on-the-fly consolidation, our concurrency protocol can support the concurrent execution of \algoname{Search}, \algoname{Insert}, and \algoname{Delete} operations.

The data structures that all queries synchronize over are the adjacency lists of the graph, the node status tracking array $H$, and a set that keeps track of all empty slots available for new graph nodes (we refer to this as the set of replaceable nodes). We use lock-free techniques on $H$ consisting of an atomic variable for each node, and lock-based synchronization for graph adjacency lists.

\myparagraph{\algoname{Delete}}
Suppose node $v$ previously represented point $x$. Since \algoname{Delete} only involves changing $H$ (\Cref{algorithm:cleann_delete}), $\algoname{Delete}(v)$ only needs to successfully perform one CAS on $v$. Any concurrent \algoname{Search} query that processes $v$ during the final neighbor selection (Line~\ref{line:cs_filter_tombstone} in~\Cref{algorithm:cleann_search}) after \algoname{Delete}($v_x$) will not include $x$ in the result.

\myparagraph{\algoname{Search}}
Following the concurrency techniques in \sysname{FreshVamana}, when the graph is traversed, adjacency lists are protected by read-write locks (shared reader and exclusive writer).

After $C$ consolidations have been performed for a deleted node $v_x$, a \algoname{Search} calls \algoname{TryMarkReplaceable} on the node (Line~\ref{line:cdbs_free} in \Cref{algorithm:clean_dynamic_beam_search}) when it encounters the node, which inserts $v$ into the set $M$ of replaceable nodes, making $v$ available for representing a new data point. Only one thread will successfully perform the CAS, guaranteeing that $v$ is released into the replaceable nodes set exactly once due to one previous \algoname{Delete}. If another thread concurrently calls \algoname{TryMarkReplaceable} on $v$ and succeeds before the current thread, there are three possible scenarios: (1) the current thread sees that $v$ is already replaceable and does not modify $H(v)$; (2) a new \algoname{Insert} and a sequence of consolidations are performed for $v$ (now representing a new point $y$), the current thread sees $H(v) < C$ and does not modify $H(v)$, leaving $v$ not replaceable as expected; and (3) a new insert and $C$ consolidations are performed for $v$ (now representing $y$), the current thread sees $H(v) = C$ and marks $v$ as replaceable, stealing work from the last thread that consolidates $v_y$. In any case, the status of $v$ is well-defined.

\myparagraph{\algoname{Insert}}
The concurrency control in $\algoname{Insert}$ queries is mostly the same as for \algoname{Search}, except for two key differences. First, in \algoname{RobustInsertData} (\Cref{algorithm:cleann_insert_data}) where a graph node is chosen to represent the data point, the node must have  replaceable status ($H(w)=-2$).
Second, when the adjacency lists of the new node and its neighbors are updated (Lines~\ref{line:robust_insert_prunes}--\ref{line:robust_insert_reverse_edges} of~\Cref{algorithm:robust_insert}), the corresponding exclusive locks of the adjacency lists are held.

\section{Experiments}\label{section:experiments}
We study the performance of \system, which uses all of the methods from Sections~\ref{section:bridge} and~\ref{section:dynamic_consolidate}, and compare with existing solutions.

\subsection{Experiment Setup}\label{subsec:experiment_setup}

\myparagraph{Sliding Window Batched Update}
Each experiment consists of multiple rounds. In each round, we insert a batch of new data points and delete an equal number of the oldest data points. The index is first constructed by inserting the first half of the entire dataset. The numbers of  inserted and deleted points in each subsequent batch are each 1\% of the current index size. Real-world datasets are streamed in the order given to reflect realistic distribution shifts where applicable. At the end of each round, we issue all search queries in the dataset, and calculate the recall (\Cref{definition:recall}). Unless otherwise specified, all \algoname{Insert} queries and a randomly selected 5\% of \algoname{Search} queries perform \sysname{GuidedBridgeBuild} (denoted $\neg \varname{performance\_sensitive}$ in~\Cref{algorithm:clean_dynamic_beam_search}). Since \system{} amortizes graph repairs over the workload, when  \system{} detects bloat above a threshold, we issue  \algoname{Search} queries for 10\% of deleted points so that more on-the-fly consolidations are performed. These \algoname{Search} queries are not counted as separate queries and their latencies are accounted for in the throughput calculation of delete batches.

\myparagraph{Sliding Window Mixed Update}
This setting runs all types of queries concurrently. Recall is not measured in this setting since the ground truth is not well-defined. 


\myparagraph{Implementations}
The implementation\footnote{Available at \texttt{https://github.com/SylviaZiyuZhang/cleanann-rust}} of \system is based on the open-source in-memory version of \sysname{DiskANN} (\sysname{Vamana})~\cite{diskann_github}. We use \texttt{Tokio} for parallelism following the \sysname{DiskANN} Rust codebase. \algoname{CleanANN-} is our system without \algoname{GuidedBridgeBuild}.



We compare with the following baselines:
\begin{itemize}[topsep=1pt,itemsep=0pt,parsep=0pt,leftmargin=10pt]
\item \sysname{IP-DiskANN}~\citep{xu2025inplaceupdatesgraphindex}. \sysname{IP-DiskANN} is implemented by the original authors and uses the same \sysname{DiskANN} codebase that we use.
\item \sysname{Wolverine++}~\cite{liu2025wolverine}. Since \sysname{Wolverine++}~\cite{liu2025wolverine} was originally implemented on \texttt{HNSW}~\cite{Malkov2025hnswlib}, we re-implemented the algorithm on the same \sysname{DiskANN} codebase.
\end{itemize}

On \algoname{Delete}$(v_x)$, both baselines mark $v_x$ as a tombstone before performing immediate graph repairs. Both baselines use global batch operations to clean up residual incoming edges for tombstone nodes. We refer readers to~\cite{xu2025inplaceupdatesgraphindex} for comparisons with \sysname{FreshVamana}.

\myparagraph{Hyperparameter Choices}
We now revisit the hyperparameters required by \system and other baselines and discuss how they are set in the experiments. $L$ configures the scope of \algoname{Search} backtracking. $L_I$ configures the scope of \algoname{Insert} backtracking. $R$ and $\alpha$ configure the sparsity of the graph via \algoname{RobustPrune}. In all experiments, we set $R=64$, $\alpha=1.2$, and $L_I=128$. We report results from varying values of $L$ to reflect the capabilities of different systems at different target recall ranges.

$\calS$ and \algoname{HeuristicPredicate} configure \sysname{GuidedBridgeBuild} in \system. We set 
$\calS = \{\log_{2}|\datasetname{D}|-1, \log_{2}|\datasetname{D}|, \log_{2}|\datasetname{D}|+1\}$, and $\algoname{HeuristicPredicate}(v_1, v_2) = \texttt{True}$ if and only if $r(v_1) = r(v_2)$ in the search tree $\calT$ of \algoname{BridgeBuilderSearch} (\Cref{algorithm:guided_bridge_building}). $C$ controls how eagerly the semi-lazy cleaning method completely cleans a tombstone. We set $C=7$ except in the microbenchmark studying the sensitivity of $C$ in~\Cref{subsec:tradeoff}.

\myparagraph{Datasets}
We evaluate on 10-million and 100-million point segments of three datasets representing real-world large-scale embedding-based information retrieval using two different distance functions ($\ell_2$ and cosine similarity) and different data distributions, with dimensionalities ranging from 100 to 512. By default, the results presented in the detailed analysis are on the 10-million point segments. The datasets are listed in Table \ref{table:datasets}. \datasetname{MS-Turing}~\cite{bab21}  and \datasetname{MS-SpaceV}~\cite{spacev1bdataset} are based on Web-scale general search queries, while \datasetname{RedCaps}~\cite{redcapsdataset, redcapsdataset_original} is based on site-level, more specialized search queries from Reddit. \datasetname{MS-Turing} and \datasetname{RedCaps} have out-of-distribution search queries. \datasetname{MS-SpaceV} and \datasetname{RedCaps} have real-world distribution shifts. 

\begin{table*}[h!]
  \caption{Dataset Information.}
  \label{table:datasets}
    \setlength{\tabcolsep}{3pt}
  \begin{tabular}{ccccclc}
    \toprule
    Name & \#Queries & Dim. & Similarity & Distribution Shift & Domain & Out-of-Distribution Queries\\        
    \midrule
    \datasetname{MS-Turing}~\cite{bab21} & 10,000 & 128 & $\ell_2$ & No & Web-scale Document Search & Yes\\
    
    \datasetname{MS-SpaceV}~\cite{spacev1bdataset} & 29,316 & 100 & $\ell_2$ & Yes & Web-scale Document Search & No\\
    \datasetname{RedCaps}~\cite{redcapsdataset} & 800  & 512 & \cosine & Yes & Text-to-Multimodal Search & Yes\\

  \bottomrule
\end{tabular}
\end{table*}

\begin{table*}[ht]
  \caption{Main Benchmark Result Summary. 
  \sysname{IP} is \sysname{IP-DiskANN}~\cite{xu2025inplaceupdatesgraphindex} and \sysname{W++} is \sysname{Wolverine++}~\cite{liu2025wolverine}.
  "$\times$ throughput" is the multiplicative speedup of throughput (measured in the Sliding Window Mixed Update setting) of \system compared to \sysname{IP-DiskANN}~\cite{xu2025inplaceupdatesgraphindex} and \sysname{Wolverine++}~\cite{liu2025wolverine}. In this table, all systems use $L=200$ and $L_I=128$. \sysname{Wolverine++}~\cite{liu2025wolverine} only supports Euclidean distance, which does not apply to \datasetname{RedCaps}.}
  \label{table:result_summary}
    \setlength{\tabcolsep}{3pt}
  \begin{tabular}{ccccccc}
    \toprule
    Name  & Size & \#Queries & \system| \sysname{IP} | \sysname{W++} recall10@10 & \system| \sysname{IP} | \sysname{W++} recall50@50 & \sysname{IP} | \sysname{W++} $\times$ throughput \\        
    \midrule
    
    \datasetname{MS-SpaceV}~\cite{spacev1bdataset} & 10M  & 29,316  & 98.11\% $\mid$ 98.11\% $\mid$ 97.68\% & 96.94\% $\mid$ 96.82\% $\mid$ 96.51\% &  1.24$\times$ $\mid$ 18.14$\times$ \\

    \datasetname{MS-SpaceV}~\cite{spacev1bdataset} & 100M  & 29,316  & 97.49\% $\mid$ 97.59\% $\mid$ 97.14\% &
    96.57\% $\mid$ 96.61\% $\mid$ 96.21\%  &  1.25$\times$ $\mid$ 16.9$\times$ \\
    
    \datasetname{MS-Turing}~\cite{bab21}  & 10M & 10,000  & 96.1\% $\mid$ 96.32\% $\mid$ 95.17\% & 92.19\% $\mid$ 92.23\% $\mid$ 91.13\% &  1.21$\times$ $\mid$ 14.76$\times$ \\

    \datasetname{MS-Turing}~\cite{bab21}  & 100M & 10,000  & 95.62\% $\mid$ 95.84\% $\mid$ 94.67\% & 92.48\% $\mid$ 92.67\% $\mid$ 91.52\% &  1.14$\times$ $\mid$ 11.75$\times$ \\

    \datasetname{RedCaps}~\cite{redcapsdataset}  & 10M & ~800  & 95.25\% $\mid$ 95.09\% $\mid$ N/A & 94.79\% $\mid$ 92.20\% $\mid$ N/A &  1.26$\times$ $\mid$ N/A \\

  \bottomrule
\end{tabular}
\end{table*}

\myparagraph{Experiment Platform and Concurrency}
All experiments are run on a 
machine with four Intel Xeon Platinum 8176 CPUs, with a total of 112 cores with 2-way hyper-threading. The machine has a total of 1.5TB DRAM. Unless otherwise specified, all systems use 64 hyper-threads throughout the benchmarks.

\subsection{Main Benchmark Results}\label{subsection:main_benchmark_results}
We report the recall for $k=10$ and $k=50$ for all the main benchmark experiments.
We summarize the recalls (in the Sliding Window Batched Update setting) and comparisons of the overall throughput results (in the Sliding Window Mixed Update setting) of \system, \sysname{IP-DiskANN}~\cite{xu2025inplaceupdatesgraphindex}, and \sysname{Wolverine++}~\cite{liu2025wolverine} in~\Cref{table:result_summary}, running both settings for 100 rounds after the index is constructed via insert-only batches (the first 100 batches). We also include results for \sysname{CleanANN-}, \sysname{CleanANN} without \algoname{GuidedBridgeBuild}.

\textit{\system{} achieves recall competitive with \sysname{IP-DiskANN}~\cite{xu2025inplaceupdatesgraphindex} and \sysname{Wolverine++}~\cite{liu2025wolverine} while delivering higher overall throughput than both}. For datasets with a real-world distribution shift and/or out-of-distribution queries, \system{} does not exhibit recall degradation.

\begin{figure}[!t]
\centering
\includegraphics[width=\linewidth]{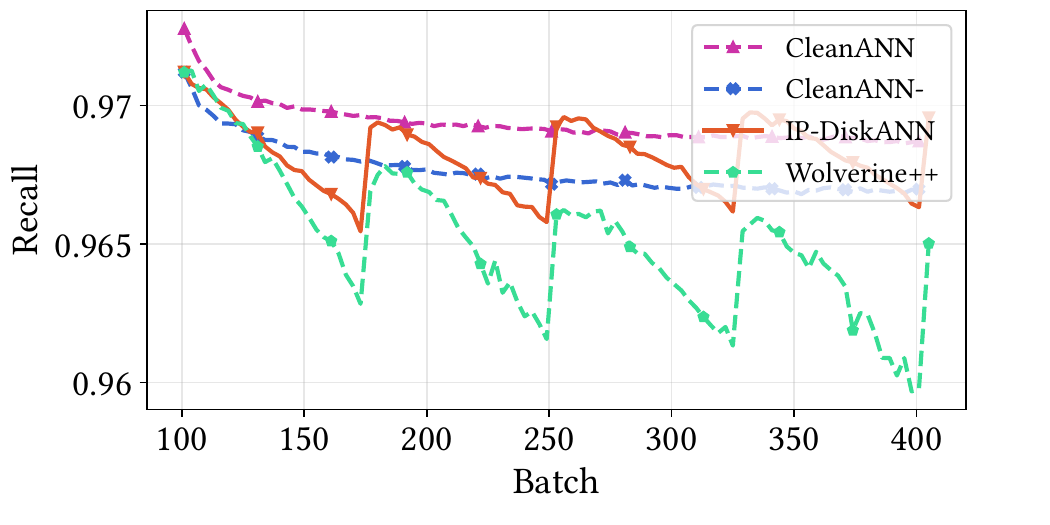} 
\caption{Recall 50@50 over time for \datasetname{MS-SpaceV} with $L=200$.}\label{figure:main_benchmark_recall_spacev}

\end{figure}

\begin{figure}[!t]
\centering
\includegraphics[width=\linewidth]{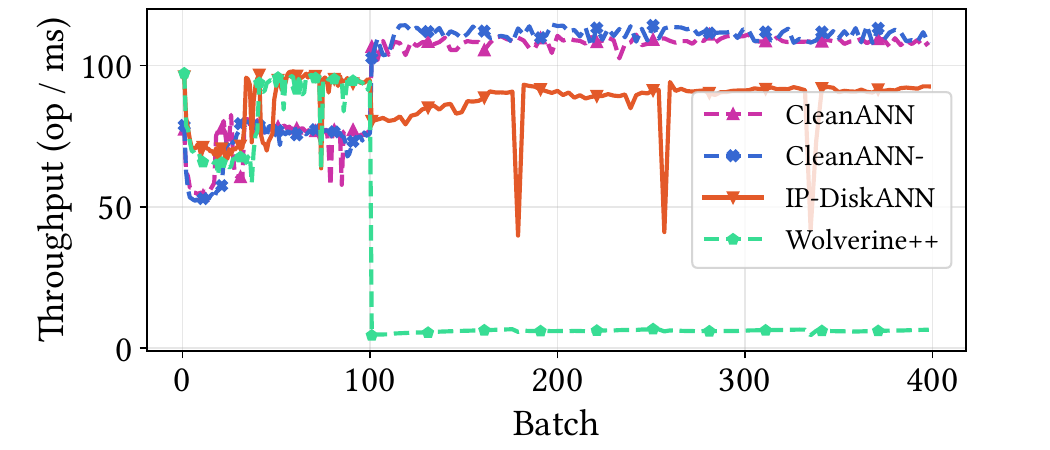}
\caption{Overall throughput for \datasetname{MS-SpaceV} in the Sliding Window Mixed Update setting.}
\label{figure:main_benchmark_throughput_spacev}

\end{figure}


\myparagraph{Search Quality}
Figure~\ref{figure:main_benchmark_recall_spacev} presents the recalls of our system and baselines on \datasetname{MS-SpaceV} in the Sliding Window Batched Update setting, showing that \emph{\system maintains high index quality when undergoing updates}. Compared to \system{}, \sysname{IP-DiskANN}~\cite{xu2025inplaceupdatesgraphindex} and \sysname{Wolverine++}~\cite{liu2025wolverine} both show recall degradation between global consolidation phases, with the degradation in \sysname{Wolverine++}~\cite{liu2025wolverine} being worse. This problem worsens as the search $k$ gets closer to $L$.
\sysname{CleanANN-} achieves $0.3\%$ lower recall due to missing the graph structure improvements from \algoname{GuidedBridgeBuild}.

\myparagraph{Efficiency}
Figure~\ref{figure:main_benchmark_throughput_spacev} shows the overall throughput of different systems MS-SpaceV with 64 hyper-threads in the Sliding Window Mixed Update setting. The first 100 batches are streaming \algoname{Inserts} and the subsequent batches are fully concurrent mixed updates. Compared to \sysname{IP-DiskANN}, which maintains a similar recall,
\system{} has 1.14--1.26$\times$ higher overall throughput and does not experience significant throughput drops due to the batch cleaning. \system{} has 12--18$\times$ higher overall throughput compared to \sysname{Wolverine++}~\cite{liu2025wolverine}. After the first 100 insert-only batches, the overall throughput of \sysname{Wolverine++} drops significantly due to the cost of graph repair candidate evaluation for \algoname{Delete}s. \sysname{CleanANN-} has slightly higher overall throughput than \sysname{CleanANN} due to the absence of the cost of \algoname{GuidedBridgeBuild}.


\subsection{Tradeoffs, Scalability, and Ablation Studies}\label{subsec:tradeoff}

\begin{figure}[!t]
\centering
\includegraphics[width=\linewidth]{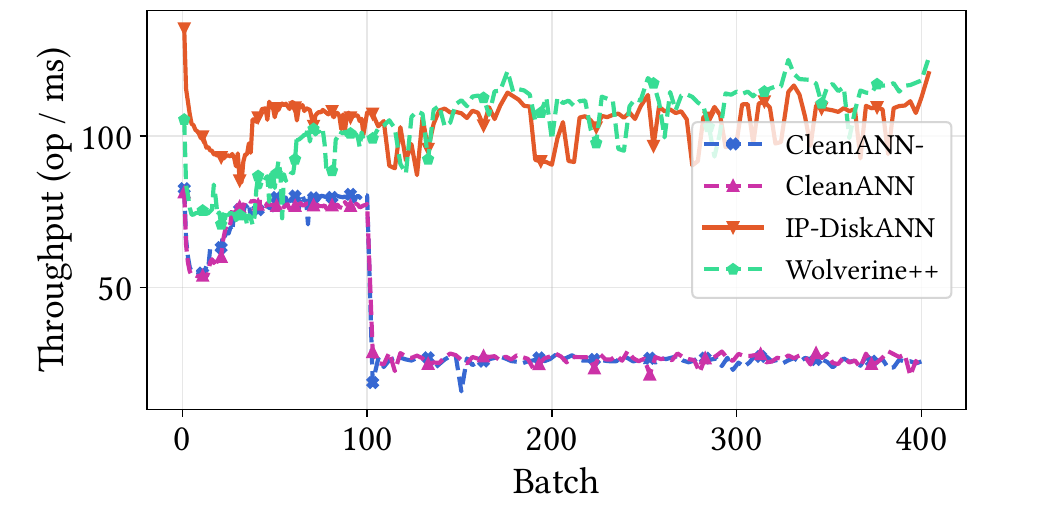} 
\caption{Insert throughput in the Sliding Window Batched Update setting on MS-SpaceV.
}\label{figure:insert_throughput_vs_batch_spacev}
\end{figure}

\begin{figure}
    \centering
    \includegraphics[width=\linewidth]{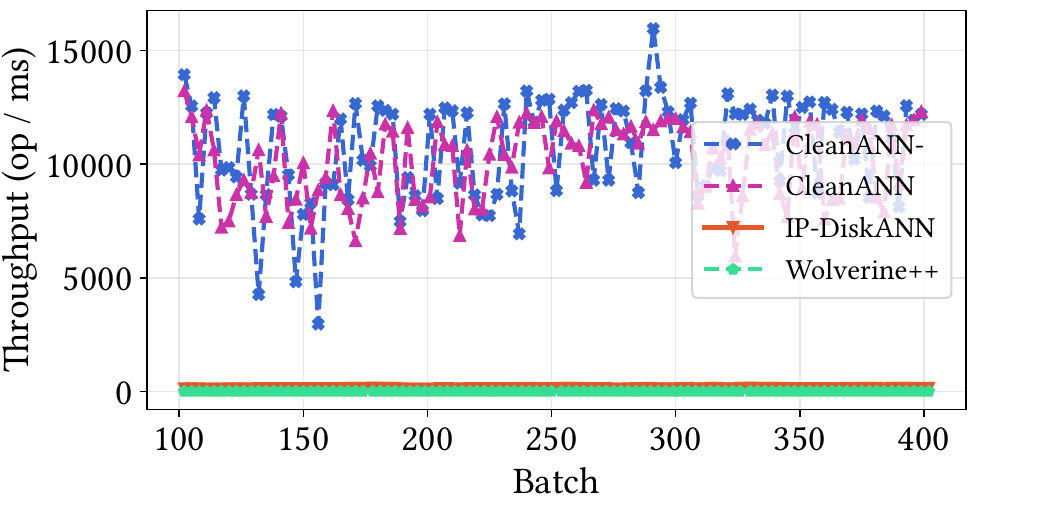}
    \caption{Delete throughput in the Sliding Window Batched Update setting on MS-SpaceV.}
    \label{figure:delete_throughput_vs_batch_spacev}
\end{figure}

\begin{figure}
    \includegraphics[width=\linewidth]{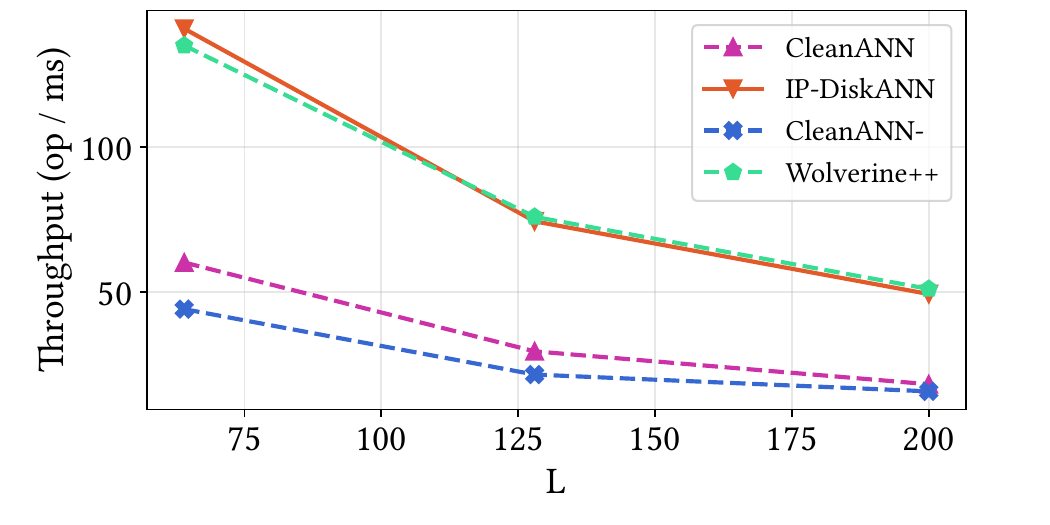}
    \caption{Search Throughput vs.\ different values of $L$ in the Sliding Window Batched Update setting on MS-SpaceV.}
    \label{figure:search_throughput_vs_l_spacev}
\end{figure}

\begin{figure}
    \includegraphics[width=\linewidth]{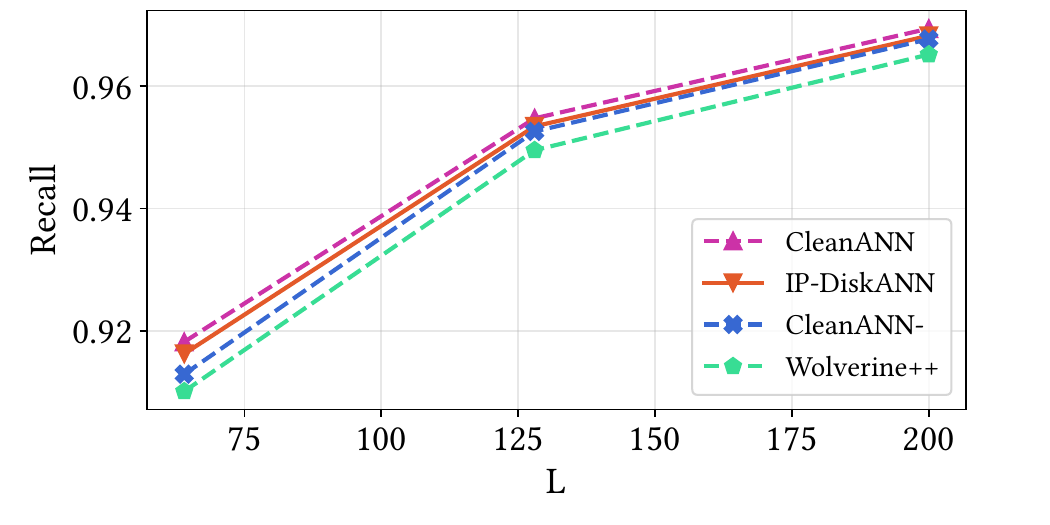}
    \caption{Recall 50@50 vs.\ different values of $L$ in the Sliding Window Batched Update setting on MS-SpaceV.}
    \label{figure:search_recall_vs_l_spacev}
\end{figure}

\myparagraph{Tradeoffs between Search/Update Throughput and Search Quality}\label{subsubsec:tradeoff}
Figures~\ref{figure:insert_throughput_vs_batch_spacev}--\ref{figure:search_throughput_vs_l_spacev} show the tradeoffs between \algoname{Search}, \algoname{Insert}, and \algoname{Delete} throughputs among different systems on MS-SpaceV measured in the Sliding Window Batched Update setting.
\system{} has lower batched \algoname{Insert} and \algoname{Search} throughputs because graph repair operations are amortized onto the \algoname{Insert} and \algoname{Search} batches. Correspondingly, the \algoname{Delete} batch throughput is much higher since~\Cref{algorithm:cleann_delete} does not perform graph updates (\Cref{figure:delete_throughput_vs_batch_spacev}). We present recalls at different values of $L$ in~\Cref{figure:search_recall_vs_l_spacev}. 

\begin{figure}[!t]
\centering
\includegraphics[width=\linewidth]{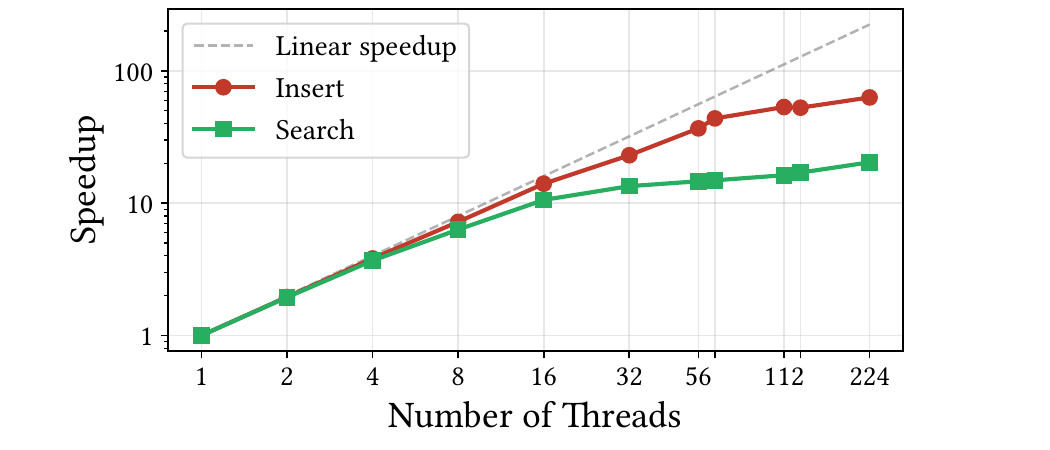}
        \caption{Parallel scaling of \system on MS-SpaceV in the Sliding Window Batched Update setting.
        }
        \label{figure:scaling-study}
\end{figure}

\myparagraph{Parallel Scalability}
We study the scalability of \system{} in the Sliding Window Batched Update setting for 100 rounds on \datasetname{MS-SpaceV} with an initial index size of 5 million using 1--224 hyper-threads.
These results are shown in ~\Cref{figure:scaling-study}. 
Compared to the single-threaded performance, the \algoname{Search}/\algoname{Insert} throughputs are 1.95$\times$/1.95$\times$ at 2 threads, 6.32$\times$/7.22$\times$ at 8 threads, 10.58$\times$/14.03$\times$ at 16 threads, 14.64$\times$/36.76$\times$ at 56 threads, 16.27$\times$/53.30$\times$ at 112 threads, and 20.36$\times$/62.98$\times$ at 224 threads. 
These findings are consistent with previous results~\cite{parlayann} on \sysname{Vamana}.
\system{} maintains at least $90\%$ recall throughout this experiment.

\myparagraph{Ablation Study for \algoname{GuidedBridgeBuild}}
By comparing \system{} and \system{}- in~\Cref{figure:main_benchmark_recall_spacev} and~\Cref{figure:main_benchmark_throughput_spacev}, we can see that \algoname{GuidedBridgeBuild} delivers a consistent recall improvement ($0.3\%$) at the cost of $3\%$ lower overall throughput. We note that in previous methods, achieving a small recall improvement in the high (95\%+) recall region often requires significantly increasing the work required~\citep{ann_benchmark}.
Furthermore, \algoname{GuidedBridgeBuild} \textit{improves the convergence of greedy searches and makes \algoname{Search} queries more efficient}. With the same backtracking budget $L$, \system{} converges faster (\Cref{figure:search_throughput_vs_l_spacev}) and to a higher recall (\Cref{figure:search_recall_vs_l_spacev}), achieving a superior throughput-recall tradeoff. 


Lastly, we compare the edges that \algoname{GuidedBridgeBuild} adds to the graph index with running index construction and \algoname{Insert} under a higher out-degree bound $R$ to demonstrate that \algoname{GuidedBridgeBuild} improves the navigability of the graph even while controlling for the number of edges.
We additionally run the experiment presented in~\Cref{table:result_summary} for MS-SpaceV on \system{} with $R = 64$ and \system{}- with $R = 74$. These two configurations achieve the same recall, with the former resulting in an average degree of $39$ and the latter $44$, suggesting that the benefits of \algoname{GuidedBridgeBuild} do not merely come from making the graph denser. 
These results suggest that \algoname{GuidedBridgeBuild} fundamentally improves the graph quality beyond \algoname{RobustPrune}.

\begin{table*}[ht]
  \caption{Recall improvement and \algoname{Insert}/\algoname{Search} efficiency compared to \system{}- for different \algoname{GuidedBridgeBuild} heuristics. $\calS$ is varied and the hyperparameter of each heuristic is fixed.}
  \label{table:gbb-heuristics-comparison}
    \setlength{\tabcolsep}{3pt}
  \begin{tabular}{cccccc}
    \toprule
    Heuristic  & Hyperparameter & + Recall at $L = 128$ & $\times$ Insert Throughput & $\times$ Search Throughput\\        
    \midrule
    
    Random Sampling  & $p=0.05$ & $0.13$--$0.7\%$  & $0.73$--$0.88\times$ & $0.84$--$1.14\times$ \\

    Distance of LCA & $l=3$ & $0.09$--$0.37\%$ & $0.95$--$1.09\times$ & $0.87$--$1.13\times$ \\

    Connecting ancestors-descendants & $l=3$ & $0.03$--$0.04\%$ & $0.91$--$1.17\times$ & $0.99$--$1.06\times$ \\

    Bounded depth-difference constraint & $\delta = 1$ & $0.15$--$0.82\%$ & $0.75$--$0.9\times$ & $1.05$--$1.5\times$ \\

    Equal depth constraint (proposed) & --- &$0.09$--$0.56\%$ & $0.88$--$1.02\times$ & $0.99$--$1.06\times$ \\
  \bottomrule
\end{tabular}
\end{table*}

\myparagraph{Choice of \algoname{GuidedBridgeBuild} Heuristics}
The \algoname{HeuristicPredicate} in the main benchmark constrains the depths of nodes to be the same. We considered several other plausible heuristics:

\begin{enumerate}[leftmargin=*]
    \item Random Sampling: Given $\calS$, for $v, w \in \{u\in\calT \mid r(u)\in\calS\}$, $\algoname{HeuristicPredicate}(v, w) = \mathbf{true}$ with probability $p$.
    \item Distance of Lowest Common Ancestor (LCA): Given $\calS$ and a threshold $l\in \mathbb{N}^+$, for $v, w\in \{u\in\calT\mid r(u)\in\calS\}$, let $u^\star$ be the LCA of $v$ and $w$. $\algoname{HeuristicPredicate}(v, w) = \mathbf{true}$ if and only if $r(v) - r(u^\star) > l$ and $r(w) - r(u^\star) > l$.
    \item Connecting ancestors-descendants: Given $\calS$ and a threshold $l\in \mathbb{N}^+$, for $v, w\in \{u\in\calT\mid r(u)\in\calS\}$, $\algoname{HeuristicPredicate}(v, w) = \mathbf{true}$ if and only if $|r(v) - r(w)| > l$ and $v$ is an ancestor or descendant of $w$.
    \item Bounded depth-difference constraint: Given $\calS$, for $v, w\in \{u\in\calT\mid r(u)\in\calS\}$, $\algoname{HeuristicPredicate}(v, w) = \mathbf{true}$ if and only if $|r(v) - r(w)| < \delta$.
\end{enumerate}

We present results for these heuristics in ~\Cref{table:gbb-heuristics-comparison}.
We find that the exact choice of heuristic does not generally produce a qualitative difference. This is likely because as long as candidate edges close the same gap among the same well-connected regions, the exact edge to add is not crucial for performance and accuracy. We suggest using the equal depth constraint because it has the most consistent performance and the smallest overhead across datasets. 
Furthermore, we found that the choice of $\calS$ is much more important than the choice of the heuristic.

\begin{figure}[!t]
\centering
\includegraphics[width=\linewidth]{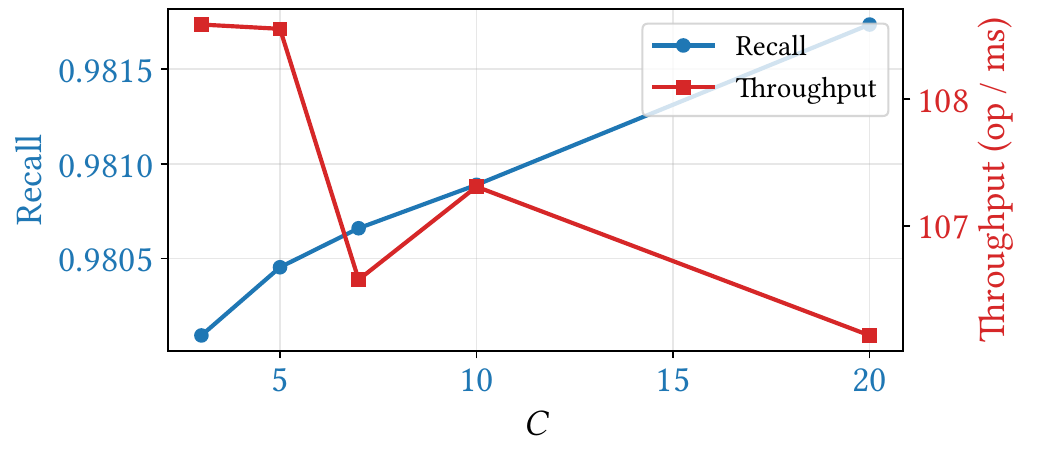} 
\caption{Average throughput and recall of \system on \datasetname{MS-SpaceV} vs.\ $C$.
}\label{figure:c_sensitivity_spacev}
\end{figure}

\myparagraph{Semi-lazy Memory Cleaning Hyperparameter Sensitivity}\label{subsubsection:c_sensitivity}
We present a microbenchmark that studies the sensitivity of the cleaning threshold hyperparameter $C$ on  MS-SpaceV (Figure~\ref{figure:c_sensitivity_spacev}). The recall increases slightly and exhibits  diminishing returns as $C$ increases. A higher value of $C$ means that more incoming edges of a tombstone need to be consolidated, causing lower throughput.

\myparagraph{Effect of Random Edges}\label{subsubsec:ablation_random_edge_contamination}
We study whether the technique of overwriting graph nodes with dangling incoming edges significantly affects the graph quality. We compare the recall and throughput under the Sliding Window Batched Update setting between the case where newly inserted data points reuse old graph nodes and the case where they do not, holding all other hyperparameters equal. The differences in these metrics are negligible. Additionally, we implemented a simple spatially aware  strategy. When a point $x$ is inserted, the index arbitrarily picks two replaceable nodes $v_{y_1}$ and $v_{y_2}$, and chooses $\argmin_{v_{y_j}}d(x, y_j)$ to represent $x$. We find that the recall and efficiency yielded by this strategy are not significantly different from using an arbitrary slot. Overall, our experiments indicate that random edges do not significantly impact the index performance.

\section{Related Work}\label{section:related}

Static ANNS indexes have been extensively studied in the literature. Previous indexing methods can be roughly categorized into three main categories of techniques: space-partition trees, locality-sensitive hashing, and graph-based indexes. Some work~\cite{lsh_apg, spann} also combines multiple techniques for improved query performance. Space-partition trees and locality-sensitive hashing are partition-based approaches and inherently struggle with the \textit{boundary issue}: if a query point is close to a boundary of different partitions, indexed points in every partition need to be compared with the query. This is especially an issue in high dimensions since the number of possible spatial partition neighbors grows with the dimensionality.

\myparagraph{Space-partition trees}
Space-partition indexes can be naturally extended to support ANNS. Given a query $q$, the index identifies the partition $\calP(q)$ that $q$ belongs to and potentially the partitions close to $\calP(q)$. The indexed points in these identified partitions are further compared with $q$ to find the nearest neighbors. Space-partition indexes that have been applied to ANNS include $k$d-trees~\cite{rs19_kdtree_nn, jsf17_dynamic_kd_tree, hna15_parallel_kdtree_nn}, R-trees~\cite{kl97_r_tree_nns, pm97_r_tree_nns}, random projection (RP)-trees~\cite{ds15_rp_tree}, spill (SP)-trees~\cite{lmyg02_investigation_spill_tree, scann, soar_2023}, Voronoi diagrams~\cite{ks04_voronoi_knn, lkxi21_hierarchical_voronoi}, clustering~\cite{spann}, and product quantization~\cite{jds11_pq}. 
These methods are also
often combined with ideas from inverted indexes for improved
performance.

\myparagraph{Locality-sensitive hashing (LSH)}
Locality-sensitive hash functions hash 
points close to each other to hash values that are the same or close to each other with high probability.
Indexed points in the same or similar hash buckets as the query point are compared with the query point. Multiple LSH functions are often used together to boost the success probability. LSH was first proposed for ANNS in~\cite{im98_lsh_stoc, gim99_lsh_vldb} and extensively studied subsequently~\cite{dks11_fast_lsh, diim04_p_dist_lsh, mnp06_lsh_lb}. LSH-based indexes have provable guarantees, but do not attain the performance of graph-based indexes in practice~\cite{ann_benchmark}.

\myparagraph{Graph-based indexes} 
In addition to the previous graph-based indexes that we discussed already~\cite{mplk14_nsw, my20_hnsw, nsg, hd16_fanng, diskann}, the recent work \sysname{eXNG}~\cite{extended_neighborhood_graph} makes the observation that connecting nodes with their intermediate near neighbors improves the graph. This observation is similar to the intuition for \algoname{GuidedBridgeBuild}. However, \sysname{eXNG} still focuses on the static setting. We refer readers to~\cite{vldb21_survey} for a more comprehensive review of graph-based indexes.


\myparagraph{Dynamic indexes}
Fully-dynamic ANNS indexes that maintain high recall under updates are less studied. 
Using different techniques for static ANNS mentioned above in the fully-dynamic setting introduces different challenges from the ones addressed in this paper for graph-based ANNS indexes. 
Naively applying static space-partition and LSH-based solutions in the dynamic setting by inserting points into or deleting points from the partition or hash bucket creates partition/bucket imbalance problems. This can degrade the efficiency of the index and may exacerbate the boundary issue discussed above, which worsens the recall.
\sysname{SPFresh}~\cite{spfresh} uses dynamic rebalancing (based on space partitioning) to solve this problem to obtain a dynamic version of the disk-oriented clustering-based index \sysname{SPANN}~\cite{spann}.
Dynamic rebalancing techniques are orthogonal to our methods, since we focus on graph-based techniques. 
A system such as \sysname{SPFresh}, which can combine a graph-based index with other algorithms, could benefit from our methods. The existing dynamic graph-based indexes \sysname{FreshDiskANN}~\cite{freshdiskann}, \sysname{DEG}~\cite{hezel2023_deg}, and \sysname{LSH-APG}~\cite{lsh_apg}, as well as graph-based entries in the NeurIPS 2023 streaming ANN competition~\cite{simhadri2024resultsbigannneurips23},
either use expensive global updates or suffer from query quality degradation. 

Recent systems \sysname{DIGRA}~\cite{jiang2025digra} and \sysname{RangePQ}~\cite{zhang2025dynamic_range_filter_anns} propose algorithms for dynamic ANNS supporting range queries, leveraging tree structures. Their update algorithms focus on maintaining the tree structures themselves. 
Our work focuses instead on an efficient and robust solution for updating the graphs in graph-based ANNS. 

The concurrent and independent work \sysname{IP-DiskANN}~\cite{xu2025inplaceupdatesgraphindex} focuses on efficiently supporting \algoname{Delete} queries in place. When a data point $p$ is deleted, \sysname{IP-DiskANN} searches for $p$ and heuristically selects in-neighbors of $v_p$ visited during the search to absorb some other nodes close to $p$ into their neighborhoods. On the other hand, \system performs consolidation in subsequent operations and is informed by the query workload, while also providing methods to improve robustness. 



\section{Conclusion}\label{sec:conclusion}

We presented 
adaptive, robust, and efficient methods 
to improve the performance of graph-based ANNS indexes under full dynamism. We designed the \system system, which implements our methods on top of 
\sysname{Vamana}. 
Our system avoids expensive global updates and adaptively improves the index.
Future research opportunities include combining our methods with other graph-based ANNS approaches, adapting machine learning techniques from learned indexes to improve performance, and theoretically analyzing the performance and robustness of dynamic ANNS.


\begin{acks} 
This work is funded by NSF awards \#CCF-1845763, \#CCF-2316235, and \#CCF-2403237, a Google Faculty Research Award, a Google Research Scholar Award, MIT Jacobs Presidential Fellowship, NSF GRFP \#DGE-2141064, and the MIT-Google Program for Computing Innovation.
\end{acks}

\bibliographystyle{ACM-Reference-Format}
\bibliography{references}







\end{document}